\documentclass[11pt]{article}
\usepackage{amsmath}
\usepackage{amssymb}
\usepackage{amsfonts}
\usepackage{epsfig}
\usepackage{mathrsfs}
\usepackage{graphicx}
\usepackage[scriptsize]{subfigure}

\newcounter{algo}
\newenvironment{algo}[2]
  {\refstepcounter{algo}\label{#2}   \begin{center}
   \begin{minipage}{0.95\textwidth}   \hrule\smallskip
   \textbf{Algorithm \thealgo: #1}
   \par\smallskip\hrule\smallskip\ignorespaces}
  {\par\smallskip\hrule
   \end{minipage}
   \end{center}
  }
\newtheorem{theorem}{Theorem}
\newtheorem{proposition}{Proposition}

\newtheorem{lemma}{Lemma}
\newtheorem{corollary}{Corollary}

\newenvironment{proof}[1][Proof]{\noindent \textbf{#1.} }{\qedsymbol}
\newcommand{\qedsymbol}{\hspace{\fill}\rule{1.5ex}{1.5ex}}
\input{tcilatex}
\def\baselinestretch{1.33}
\def\b0{\mbox{\boldmath $0$}}

\begin{document}

\title{\vspace{-1.8cm} {\Huge Optimal Linear Precoding Strategies for
Wideband Non-Cooperative Systems based on Game Theory-Part II: Algorithms}}
\author{Gesualdo Scutari$^{1}$, Daniel P. Palomar$^{2}$, and Sergio
Barbarossa$^{1}$ \\
{\small E-mail: $\{$scutari,sergio$\}$@infocom.uniroma1.it,
palomar@ust.hk}\\
$^{1\text{ }}${\small Dpt. INFOCOM, Univ. of Rome \textquotedblleft La
Sapienza\textquotedblright, Via Eudossiana 18, 00184 Rome, Italy} \\
$^{2}$ {\small Dpt. of Electronic and Computer Eng., Hong Kong Univ. of
Science and Technology, Hong Kong.}}
\date{{\small Submitted to IEEE \textit{Transactions on Signal Processing},
February 22, 2006}\\
{\small Revised March 26, 2007. Accepted June 5, 2007.\thanks{%
This work was supported by the SURFACE project funded by the European
Community under Contract IST-4-027187-STP-SURFACE.}}\vspace{-0.9cm}}
\maketitle
\begin{abstract}
In this two-part paper, we address the problem of finding the optimal
precoding/multiplexing scheme for a set of non-cooperative links sharing the
same physical resources, e.g., time and bandwidth. We consider two
alternative optimization problems: P.1) the maximization of mutual
information on each link, given constraints on the transmit power and
spectral mask; and P.2) the maximization of the transmission rate on each
link, using finite order constellations, under the same constraints as in
P.1, plus a constraint on the maximum average error probability on each
link. Aiming at finding decentralized strategies, we adopted as optimality
criterion the achievement of a Nash equilibrium and thus we formulated both
problems P.1 and P.2 as strategic noncooperative (matrix-valued) games. In
Part I of this two-part paper, after deriving the optimal structure of the
linear transceivers for both games, we provided a unified set of sufficient
conditions that guarantee the uniqueness of the Nash equilibrium. In this
Part II, we focus on the achievement of the equilibrium and propose
alternative distributed iterative algorithms that solve both games.
Specifically, the new proposed algorithms are the following: 1) the $%
sequential$ and $simultaneous$ iterative waterfilling based algorithms,
incorporating spectral mask constraints; 2) the $sequential$ and $%
simultaneous$ gradient projection based algorithms, establishing an
interesting link with variational inequality problems. Our main contribution
is to provide sufficient conditions for the $global$ convergence of all the
proposed algorithms which, although derived under stronger constraints,
incorporating for example spectral mask constraints, have a broader validity
than the convergence conditions known in the current literature for the
sequential iterative waterfilling algorithm.
\end{abstract}

\section{Introduction and Motivation}

The goal of this two-part paper is to find the optimal
precoding/multiplexing schemes for a set of non-cooperative links sharing
the same physical resources, e.g. time and bandwidth. Two alternative
optimization problems are considered \cite{Scutari-Part I}: P.1) the
maximization of mutual information on each link, given constraints on the
transmit power and spectral emission mask, imposed by radio spectrum
regulatory bodies; and P.2) the maximization of the transmission rate on
each link, using finite order constellations, under the same constraints as
in P.1, plus a constraint on the maximum average error probability on each
link. We focus on decentralized strategies to avoid coordination among the
separated links and the heavy signaling required by a global controller that
would need to collect all relevant information from all the users. The
search for decentralized solutions motivated our formulation within the
convenient framework of game theory. We thus adopt as optimality criterion
the achievement of a Nash Equilibrium (NE) \cite{Osborne} and we cast both
optimization problems P.1 and P.2 as strategic non-cooperative
(matrix-valued) games \cite{Scutari-Part I}, where the goal of each user is
to optimize its own precoding/multiplexing matrix. In Part I of this
two-part paper \cite[Theorem 1]{Scutari-Part I}, we proved that there is no
performance loss in reducing both original matrix-valued games into a
unified vector-valued power control game, where the optimal strategy of each
user corresponds to finding the power allocation that maximizes its own
(information) rate, treating the multiuser interference due to the other
users as additive colored noise. We will refer to this power control game as
\emph{rate-maximization} game. In Part I, we proved that the solution set of
the rate-maximization game is always nonempty and derived (sufficient)
conditions that guarantee the uniqueness of the NE \cite[Theorem 2]%
{Scutari-Part I}. In this Part II, we propose alternative algorithms that
reach the Nash equilibria of the unified vector game, in a totally
distributed manner.

All the distributed algorithms used to compute the Nash equilibria of a
(rational \cite{Osborne}) strategic non-cooperative game are based on a
simple idea: Each player optimizes iteratively its own payoff function
following a prescribed updating schedule, for example, simultaneously with
the other users (i.e., according to a Jacobi scheme \cite{Bertsekas
Book-Parallel-Comp}), or sequentially (i.e., according to a Gauss-Seidel
scheme \cite{Bertsekas Book-Parallel-Comp}). Differently from the
optimization of a single-user system, where the optimal transceiver
structure can be obtained in a single shot (depending on the interference
scenario \cite{Barbarossa}-\cite{Palomar-Barbarossa}), in a competitive
multiuser context like a game, it is necessary to adopt an iterative
algorithm, as each user's choice affects the interference perceived by the
other users. However, the competitive nature of the multiuser system does
not guarantee in general the convergence of such an iterative scheme, even
if the payoff function of each player is strictly concave (or strictly
convex) in its own strategies and the NE is unique. This issue motivated
several works in the literature \cite{Mitra}-\cite{Sung-Leung}, \cite{Yu}-%
\cite{Scutari_ISIT06}, where alternative approaches have been proposed to
study the convergence of iterative algorithms in strategic non-cooperative
games.

A traditional approach comes from classical \emph{scalar }power control
problems in \emph{flat-fading }CDMA (or TDMA/FDMA) wireless networks (either
cellular or ad-hoc) \cite{Mitra}-\cite{Alpcan_et_al},\footnote{%
Note that, even though some of these papers do not contain any explicit
reference to game theory, the problems therein can be naturally reformulated
as a strategic non-cooperative game, where the Nash equilibria are the
fixed-points of proper best response correspondences.} where each user has
only one variable to optimize: its transmit power. This kind of problems can
be elegantly recast as convex optimization problems (see, e.g., \cite%
{Chiang-bookchp}-\cite{Palomar-Chiang}) or as the so called
\textquotedblleft standard\textquotedblright\ problems (in the sense of \cite%
{Yates-Jsac}-\cite{Sung-Leung}), for which distributed (either synchronous or
asynchronous) algorithms along with their convergence properties are
available \cite{Mitra}-\cite{Alpcan_et_al}, \cite{Palomar-Chiang}. The
rate-maximization game proposed in this paper is more involved, as it falls
in the class of \emph{vector }power control problems, where each player has
a vector to optimize (i.e., its power allocation across frequency bins) and the best-response
function of each user (the waterfilling mapping) is not a standard function
(in the sense of \cite{Yates-Jsac}-\cite{Sung-Leung}). Hence, the classical
framework of \cite{Mitra}-\cite{Sung-Leung} cannot be successfully applied
to our game theoretical formulation.

A special case of the rate-maximization game proposed in this paper was
studied in \cite{Yu} in the absence of spectral mask constraints, where the
authors formulated the vector power control problem for a Digital Subscriber
Line (DSL) system, modelled as a Gaussian frequency-selective interference
channel, as a two-person strategic non-cooperative game. To reach the Nash
equilibria of the game, the authors proposed the \emph{sequential }Iterative
Waterfilling Algorithm (IWFA), which is an instance of the
Gauss-Seidel scheme \cite{Bertsekas Book-Parallel-Comp}: The users maximize
their own information rates sequentially (one after the other), according to
a fixed updating order. Each user performs the single-user waterfilling
solution given the interference generated by the others as additive
(colored) noise. The most appealing features of the sequential IWFA are its
low-complexity and its distributed nature. In fact, to compute the
waterfilling solution, each user only needs to measure the
noise-plus-interference Power Spectral Density (PSD), without requiring
specific knowledge of the power allocations and the channel transfer
functions of the other users. The convergence of the sequential IWFA has
been studied in a number of works \cite{ChungISIT03}-\cite{Scutari_Thesis},
each time obtaining milder conditions that guarantee convergence. However,
despite its appealing properties, the sequential IWFA suffers from slow
convergence if the number of users in the network is large, because of the
sequential updating strategy. In addition, the algorithm requires some form
of central scheduling to determine the order in which users update their strategy.

The original contributions of this paper with respect to the current
literature on vector games \cite{Yu}-\cite{Tse} are listed next. First, to
compute the Nash equilibria of both games P.1 and P.2 (introduced in Part I
\cite{Scutari-Part I}), we generalize the sequential IWFA of \cite{Yu},
including the spectral mask constraint and a possible memory in the updating
process. Then, to overcome the potential slow convergence rate of the
sequential IWFA, we propose a new iterative algorithm, called \emph{%
simultaneous }IWFA. The simultaneous IWFA is an instance of the Jacobi
scheme \cite{Bertsekas Book-Parallel-Comp}: At each iteration, all users
update their own strategies \emph{simultaneously}, still according to the
single-user waterfilling solution, but using the interference generated by
the others in the \emph{previous} iteration. We provide results on the
convergence speed of both algorithms, showing that the simultaneous IWFA is
faster than the sequential IWFA, still keeping the desired properties of the
sequential IWFA, i.e., its distributed nature and low complexity. The second
important contribution of the paper is to provide a unified set of
sufficient conditions ensuring the \emph{global }convergence of both
algorithms. Our conditions are proved to have broader validity than those
given in \cite{Yu}-\cite{Scutari-Barbarossa-SPAWC03}, \cite{Tse} (obtained
without mask constraints) and, more recently, in \cite{Luo-Pang} (obtained
including mask constraints) for the sequential IWFA. Moreover, they show that
the range of applicability with guaranteed convergence of both sequential
and simultaneous IWFAs includes scenarios where the interfering users may be
rather close to each other. Finally, exploring the link between the Nash
equilibria of our game theoretical formulation and the solutions to the
so-called variational inequality problems \cite{Facchinei}-\cite{Nagurney},
we propose, as alternative to the IWFAs, two novel gradient projection based
iterative algorithms, namely the \emph{sequential }and \emph{simultaneous }%
Iterative\emph{\ }Gradient Projection Algorithms (IGPAs) and provide
conditions for their global convergence.

Throughout the paper, there is a common thread relating the algorithms and
the derivation of their convergence conditions: The interpretation of the
waterfilling operator as the Euclidean projector of a vector onto a convex
set. In the single-user case, this provides an alternative perspective of
the well-known waterfilling solution, that dates back to Shannon in 1949
\cite{Shannon}. Interestingly, in the multiuser case, this interpretation
plays a key role in proving the convergence of the proposed algorithms.

The paper is organized as follows. After briefly reviewing, in Section \ref%
{System Model Section}, the game theoretic formulation addressed in Part I
of the paper \cite[Theorem 2]{Scutari-Part I}, Section \ref{Sec:WF_Projector}
provides the interpretation of the waterfilling operator as a projector.
Section \ref{Section_Distributed Algorithms for Nash Equilibria} contains
the main contribution of the paper: A variety of distributed algorithms for
the computation of the Nash equilibria of the game, along with their
convergence properties. Finally, in Section \ref{Conclusions}, some
conclusions are drawn. Preliminary versions of this paper appeared in \cite%
{Scutari-Barbarossa-SPAWC03, Scutari_Thesis, Scutari_ISIT06}.\vspace{-0.2cm}

\section{System Model and Problem Formulation\label{System Model Section}}

We consider a Gaussian vector interference channel \cite{Cover}, composed by
$Q$ non-cooperative links. Aiming at finding distributed algorithms, we
focus on transmission techniques where no interference cancellation is
performed and multiuser interference is treated as additive colored noise.
Moreover, we consider a block transmission without loss of generality
(w.l.o.g.), as it is a capacity-lossless strategy for sufficiently large
block length \cite{Hir88}-\cite{Tse-book}. Then, under assumptions detailed
in Part I \cite{Scutari-Part I}, the system design consists in finding the
optimal transmit/receive matrix pair for each link independently of the
others, according to some performance metrics. In Part I of this paper \cite%
{Scutari-Part I}, we assumed as optimality criterion the achievement of the
NE and considered the two following strategic non-cooperative games:

\begin{description}
\item[P.1] The maximization of mutual information on each link, given
constraints on the transmit power and on the spectral radiation mask;\vspace{%
-0.2cm}

\item[P.2] The maximization of the transmission rate on each link, using
finite order constellations, under the same constraints as in \textbf{P.1}
plus a constraint on the average (uncoded) error probability.
\end{description}

After showing that the solution set of both games is always nonempty, 
in \cite[Theorem 1]{Scutari-Part I} we proved
that the optimal transmission strategy for each link leads to Gaussian
signaling plus the diagonal transmission through the channel eigenmodes
(i.e., the frequency subchannels), irrespective of the channel state, power
budget, spectral mask constraints and interference levels. Thanks to this
result, both \emph{matrix-valued} games \textbf{P.1 }and \textbf{P.2 }can be
recast, with no performance loss, as the following simpler \emph{vector}
power control game \cite[Theorem 1]{Scutari-Part I}:
\begin{equation}
\left(
\mathscr{G}%
\right) :\qquad \qquad \qquad
\begin{array}{l}
\limfunc{maximize}\limits_{\mathbf{p}_{q}}\quad \ R_{q}(\mathbf{p}_{q},%
\mathbf{p}_{-q}) \\
\limfunc{subject}\text{ }\limfunc{to}\text{\ \ \ }\mathbf{p}_{q}\in {%
{\mathscr{P}}%
}_{q}%
\end{array}%
,\qquad \forall q\in \Omega ,\qquad  \label{Rate Game}
\end{equation}%
where $\Omega
\triangleq%
\left\{ 1,2,\ldots ,Q\right\} $ is the set of players (i.e., active links), $%
{\mathscr{P}}_{q}$ is the set of admissible strategies of player $q:$%
\footnote{%
In order to avoid the trivial solution $p_{q}^{\star }(k)=p_{q}^{\max }(k)$
for all $k\in \{1,\ldots ,N\}$, $(1/N)\sum_{k=1}^{N}p_{q}^{\max }({k})>1$ is
assumed for all $q\in \Omega $. Furthermore, in the feasible strategy set of
each player, we can replace, w.l.o.g., the original \textit{inequality}
power constraint $(1/N)\ \sum_{k=1}^{N}p_{q}(k)\leq 1,$ with equality$,$
since, at the optimum, this constraint must be satisfied with equality from
all users.}%
\begin{equation}
{\mathscr{P}}_{q}\triangleq \left\{ \mathbf{p}_{q}\in \mathcal{\ \mathbb{R}}%
^{N}:\dfrac{1}{N}\ \sum_{k=1}^{N}p_{q}(k)=1,\text{ }0\leq p_{q}(k)\leq
p_{q}^{\max }(k),\text{ \ }\forall k\in \{1,\ldots ,N\}\right\} ,
\label{admissible strategy set}
\end{equation}%
where $p_{q}^{\max }(k)\triangleq \overline{p}_{q}^{\max }(k)/P_{q},$ with $%
\overline{p}_{q}^{\max }(k)$ denoting the maximum power that is allowed to
be allocated on the $k$-th frequency bin from the $q$-th user, and$\ R_{q}(%
\mathbf{p}_{q},\mathbf{p}_{-q})$ is the payoff function of player $q:$%
\begin{equation}
R_{q}(\mathbf{p}_{q},\mathbf{p}_{-q})=\dfrac{1}{N}\dsum\limits_{k=1}^{N}\log
\left( 1+\dfrac{1}{\Gamma _{q}}\text{ }%
\mathsf{sinr}%
_{q}(k)\right) ,  \label{Rate}
\end{equation}%
with
\begin{equation}
\mathsf{sinr}%
_{q}(k)=\frac{P_{q}\left\vert \bar{H}_{qq}(k)\right\vert
^{2}p_{q}(k)/d_{qq}^{\gamma }}{\sigma _{q}^{2}+\sum_{\,r\neq
q}P_{r}\left\vert \bar{H}_{rq}(k)\right\vert ^{2}p_{r}(k)/d_{rq}^{\gamma }}%
\triangleq \frac{\left\vert H_{qq}(k)\right\vert ^{2}p_{q}(k)}{%
1+\sum_{\,r\neq q}\left\vert H_{rq}(k)\right\vert ^{2}p_{r}(k)},
\label{SINR}
\end{equation}%
where $H_{rq}(k)\triangleq \bar{H}_{rq}(k)\sqrt{P_{r}/\left( \sigma _{q}^{2}%
\text{ }d_{rq}^{\gamma }\right) };$ $\bar{H}_{rq}(k)\ $denotes the
frequency-response on the subcarrier $k$ of the channel between source $r$
and destination $q,$ $d_{rq}$ is the distance between source $r$ and
destination $q$, and $\gamma $ is the path loss. The SNR gap $\Gamma _{q}$
in (\ref{Rate}) is set equal to $1$ if game in \textbf{P.1 }is considered,
whereas $\Gamma _{q}=(\mathcal{Q}^{-1}(P_{e,q}^{\star }/4))^{2}/3$ \cite%
{Goldsmith-Chua}, if we consider \textbf{P.2}, where $\mathcal{Q(\cdot )}$
denotes the $\mathcal{Q}$-function \cite{Proakis} and $P_{e,q}^{\star }$ is
the maximum tolerable (uncoded) average symbol error probability on link $q$.

In \cite[Theorem 2]{Scutari-Part I}, we showed that the solution set of ${%
{\mathscr{G}}%
}$ is always nonempty and coincides with the solution set of the following
nonlinear fixed-point equation:
\begin{equation}
\begin{array}{c}
\mathbf{p}_{q}^{\star }=%
\mathsf{WF}%
_{q}\left( \mathbf{p}_{1}^{\star },\ldots ,\mathbf{p}_{q-1}^{\star },\mathbf{%
p}_{q+1}^{\star },\ldots ,\mathbf{p}_{Q}^{\star }\right) =%
\mathsf{WF}%
_{q}(\mathbf{p}_{-q}^{\star })%
\end{array}%
,\quad \forall q\in \Omega ,  \label{sym_WF-sistem}
\end{equation}%
with the waterfilling operator $%
\mathsf{WF}%
_{q}\left( \mathbf{\cdot }\right) $ defined as
\begin{equation}
\left[
\mathsf{WF}%
_{q}\left( \mathbf{p}_{-q}\right) \right] _{k}%
\triangleq%
\left[ \mu _{q}-\Gamma _{q}\dfrac{1+\sum_{\,r\neq q}\left\vert
H_{rq}(k)\right\vert ^{2}p_{r}(k)}{\left\vert H_{qq}(k)\right\vert ^{2}}%
\right] _{0}^{p_{q}^{\max }(k)},\quad k\in \{1,\ldots ,N\},  \label{WF_mask}
\end{equation}%
where $\left[ x\right] _{a}^{b}$ denotes the Euclidean projection of $x$
onto the interval $[a,b].$\footnote{%
The Euclidean projection $\left[ x\right] _{a}^{b}$ \ is defined as follows:
$\left[ x\right] _{a}^{b}=a$, if $x\leq a$, $\left[ x\right] _{a}^{b}=x$, if
$a<x<b$, and $\left[ x\right] _{a}^{b}=b$, if $x\geq b$.} The water-level $%
\mu _{q}$ is chosen to satisfy the power constraint $(1/N)%
\sum_{k=1}^{N}p_{q}^{\star }(k)=1.$

Observe that system (\ref{sym_WF-sistem}) contains, as special cases, the
solutions to power control games already studied in the literature \cite{Yu}-%
\cite{Tse}, when all the players are assumed to transmit with the same power
and no spectral mask constraints are imposed (i.e., when $p_{q}^{\max
}(k)=+\infty $, $\forall q,$ $\forall k$). In this case, the Nash equilibria
of game ${%
\mathscr{G}%
}$\ are given by the classical simultaneous waterfilling solutions \cite{Yu}-%
\cite{Tse}, where $%
\mathsf{WF}%
_{q}\left( \mathbf{\cdot }\right) $ in (\ref{sym_WF-sistem}) is still
obtained from (\ref{WF_mask}) simply setting $p_{q}^{\max }(k)=+\infty $, $%
\forall q,$ $\forall k.$ However, in the presence of spectral mask
constraints, the results of \cite{Yu}-\cite{Tse} cannot be applied to system
(\ref{sym_WF-sistem}). In Part I\ of this paper \cite[Theorem 2]%
{Scutari-Part I}, we studied system (\ref{sym_WF-sistem}) and provided
sufficient conditions for the uniqueness of the solution. The problem we address here is
how to reach solutions to (\ref{sym_WF-sistem}) (leading to the Nash equilibria of $\mathscr{G}$) by means of totally distributed algorithms.\vspace{-0.2cm}

\section{Waterfilling Operator as a Projector}

\label{Sec:WF_Projector}In this section we provide an interpretation of the
waterfilling operator as a proper Euclidean projector. This interpretation
will be instrumental to prove the convergence properties of some of the
algorithms proposed in the subsequent sections.

\subsection{A new look at the single-user waterfilling solution\label%
{Sec:single-user WF}}

Consider a parallel additive colored Gaussian noise channel composed of $N$
subchannels with coefficients $\{H(k)\},$ subject to some spectral mask
constraints $\{p^{\max }(k)\}_{k=1}^{N}$ and to a global average transmit
power constraint across the subchannels. It is well-known that the
capacity-achieving solution for this channel is obtained using independent
Gaussian signaling across the subchannels with the following waterfilling
power allocation \cite{Baccarelli-Fasano-Biagi}
\begin{equation}
p^{\star }(k)=\left[ \mu -\frac{\sigma _{k}^{2}}{|H(k)|^{2}}\right]
_{0}^{p^{\max }(k)},\quad k\in \{1,\ldots ,N\},  \label{WF_mask-sinlge-user}
\end{equation}%
where $\sigma _{k}^{2}$ denotes the noise variance on the $k$-th subchannel,
$p^{\star }(k)$ is the optimal power allocation over the $k$-th subchannel$.$
The water-level $\mu $ in (\ref{WF_mask-sinlge-user}) is chosen in order to
satisfy the power constraint $(1/N)\sum_{k=1}^{N}p^{\star }(k)=1.$

We show now that, interestingly, the solution in (\ref{WF_mask-sinlge-user})
can be interpreted as the Euclidean projection of the vector $-\boldsymbol{%
\mathsf{insr}},$ defined as%
\begin{equation}
\boldsymbol{\mathsf{insr}}%
\triangleq%
\left[ \sigma _{1}^{2}/|H(1)|^{2},\ldots ,\sigma _{N}^{2}/|H(N)|^{2}\right]
^{T}  \label{insr}
\end{equation}%
onto the simplex
\begin{equation}
\mathcal{S}%
\triangleq%
\left\{ \mathbf{x}\in \mathcal{\ \mathbb{R}}^{N}:\frac{1}{N}\
\sum_{k=1}^{N}x_{k}=1,\text{ }0\leq x_{k}\mathbf{\leq }p^{\max }(k),\quad
\forall k\in \{1,\ldots ,N\}\right\} .  \label{simplex X}
\end{equation}

\begin{lemma}
\label{Euclidean Projection} The Euclidean projection of the $N$-dimensional
real nonpositive vector $-\mathbf{x}_{0}%
\triangleq%
-[x_{0,1},\ldots ,x_{0,N}]^{T}$ onto the simplex $\mathcal{S}$ defined in (%
\ref{simplex X}), denoted by $\left[ -\mathbf{x}_{0}\right] _{\mathcal{S}}$,
is by definition the solution to the following convex optimization problem:
\begin{equation}
\begin{array}{ll}
\limfunc{minimize}\limits_{\mathbf{x}} & \left\Vert \mathbf{x-(-\mathbf{x}%
_{0})}\right\Vert _{2}^{2}\smallskip \\
\limfunc{subject}\text{ }\limfunc{to} & 0\leq x_{k}{\leq }p^{\max }(k)\text{,%
}\quad \forall k\in \{1,\ldots ,N\} \\
& \dfrac{1}{N}\ \dsum\limits_{k=1}^{N}x_{k}=1.%
\end{array}
\label{convex-projection}
\end{equation}%
and assumes the following form:
\begin{equation}
x_{k}^{\star }=\left[ \mu -x_{0,k}\right] _{0}^{p^{\max }(k)},\quad k\in
\{1,\ldots ,N\},\vspace{-0.3cm}  \label{optimal solution}
\end{equation}%
where $\mu >0$ is chosen in order to satisfy the constraint $\left( 1/N\right) \
\sum_{k=1}^{N}x_{k}^{\star }=1.$
\end{lemma}

\begin{proof}
See Appendix \ref{proof_of_Euclidean Projection}.
\end{proof}

Lemma \ref{Euclidean Projection} is an extension of \cite[Lemma 1]%
{Daniel-paper} to the case where interval bounds $[0,$ $p^{\max }(k)]$ are
included in the optimization. But what is important to remark about Lemma %
\ref{Euclidean Projection} (and this is a contribution of this paper) is
that it allows us to interpret the waterfilling operator as a projector,
according to the following corollary.

\begin{corollary}
\label{Corollary-WF-project}The waterfilling solution $\mathbf{p}^{\star }=%
\mathbf{[}p^{\star }(1),\ldots ,p^{\star }(N)\mathbf{]}^{T}$ in (\ref%
{WF_mask-sinlge-user}) can be expressed as the projection of $-\boldsymbol{%
\mathsf{insr}}$ given in (\ref{insr}) onto the simplex $\mathcal{S}$ in (\ref%
{simplex X}):
\begin{equation}
\mathbf{p}^{\star }=\left[ -\boldsymbol{\mathsf{insr}}\right] _{\mathcal{S}}.
\label{WF-op}
\end{equation}
\end{corollary}

\begin{corollary}
The waterfilling solution in the form%
\begin{equation}
p^{\star }(k)=\left[ \frac{\mu }{w_{k}}-\frac{\sigma _{k}^{2}}{|H(k)|^{2}}%
\right] _{0}^{p^{\max }(k)},\qquad k\in \{1,\ldots ,N\},
\label{WF-Single-user}
\end{equation}%
where $\mathbf{w}=\mathbf{[}w_{1},\ldots ,w_{N}\mathbf{]}^{T}$ is any
positive vector, can be expressed as the projection with respect to the
weighted Euclidean norm\footnote{%
The weighted Euclidean norm $\left\Vert \mathbf{x}\right\Vert _{2,\mathbf{w}%
} $ is defined as $\left\Vert \mathbf{x}\right\Vert _{2,\mathbf{w}}%
\triangleq%
(\sum\nolimits_{i}w_{i}\left\vert x_{i}\right\vert ^{2})^{1/2}$ \cite{Horn85}%
.} with weights $w_{1},\ldots ,w_{N},$ of $-\boldsymbol{\mathsf{insr}}$
given in (\ref{insr}) onto the simplex $\mathcal{S}$ in (\ref{simplex X}):
\begin{equation}
\mathbf{p}^{\star }=\left[ -\boldsymbol{\mathsf{insr}}\right] _{\mathcal{S}%
}^{\mathbf{w}}.  \label{WF-proj}
\end{equation}
\end{corollary}

The graphical interpretation of the waterfilling solution as a Euclidean
projector, for the single-carrier two-user case, is given in Figure \ref%
{Figure_WF_project}: For any $\boldsymbol{\mathrm{insr}}\equiv (\mathrm{{insr%
}_{1},{insr}_{2})}$ corresponding to a point in the interior of the gray
region (e.g., point $A$), the waterfilling solution allocates power over
both the channels. If, instead, the vector $\boldsymbol{\mathrm{insr}}$ is
outside the gray region (e.g., point $B$), all the power is allocated only
over one channel, the one with the highest normalized gain.

\begin{figure}[h]
\par
\hspace{2.9cm} \includegraphics[height=17cm, width=12cm]{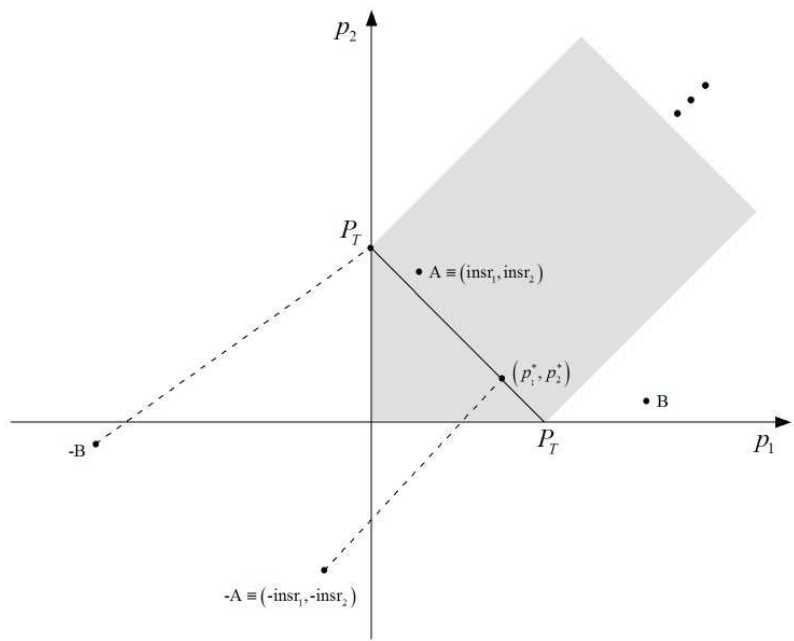}
\vspace{-8.9cm}
\caption{{\protect\footnotesize Graphical interpretation of waterfilling
solution (\protect\ref{WF_mask-sinlge-user}) as a projection onto the
two-dimensional simplex.}}
\label{Figure_WF_project}
\end{figure}

\subsection{Simultaneous multiuser waterfilling\label{Sec:Nash Equilibria}}

In the multiuser scenario described in game ${%
\mathscr{G}%
}$, the optimal power allocation of each user depends on the power
allocation of the other users through the received interference, according
to the simultaneous multiuser waterfilling solution in (\ref{sym_WF-sistem}).

As in the single-user case, introducing the vector $%
\boldsymbol{\mathsf{insr}}%
_{q}(\mathbf{p}_{-q}),$ defined as%
\begin{equation}
\left[
\boldsymbol{\mathsf{insr}}%
_{q}(\mathbf{p}_{-q})\right] _{k}%
\triangleq%
\Gamma _{q}\frac{1+\sum_{\,r\neq q}\left\vert H_{rq}(k)\right\vert
^{2}p_{r}(k)}{\left\vert H_{qq}(k)\right\vert ^{2}},\qquad k\in \{1,\ldots
,N\},  \label{INR_q}
\end{equation}%
and invoking Lemma \ref{Euclidean Projection}, we obtain the following
corollary.

\begin{corollary}
\label{Corollary-Multiuser_WF_Projection}The waterfilling operator ${\mathsf{%
WF}}_{q}\left( \mathbf{p}_{-q}\right) $ in (\ref{WF_mask}) can be expressed
as the projection of $-\boldsymbol{\mathsf{insr}}_{q}(\mathbf{p}_{-q})$
defined in (\ref{INR_q}) onto the simplex ${\mathscr{P}}_{q}$ given in (\ref%
{admissible strategy set}):
\begin{equation}
{\mathsf{WF}}_{q}\left( \mathbf{p}_{-q}\right) =\left[ -\boldsymbol{\mathsf{%
insr}}_{q}(\mathbf{p}_{-q})\right] _{{{\mathscr{P}}}_{q}}.
\label{WF-projection}
\end{equation}
\end{corollary}

Comparing (\ref{sym_WF-sistem}) with (\ref{WF-projection}), it is
straightforward to see that all the Nash equilibria of game ${%
\mathscr{G}%
}$ can be alternatively obtained as the fixed-points of the mapping defined
in (\ref{WF-projection}), whose existence is guaranteed by \cite[Theorem 2]%
{Scutari-Part I}:%
\begin{equation}
\mathbf{p}_{q}^{\star }=\left[ -\boldsymbol{\mathsf{insr}}_{q}(\mathbf{p}%
_{-q}^{\star })\right] _{{{\mathscr{P}}}_{q}},\quad \forall q\in \Omega .
\label{fixed-point_NE}
\end{equation}%
In Appendix \ref{Appendix_WF_project}, we provide the key properties of the
mapping in (\ref{WF-projection}), that will be instrumental to obtain
sufficient conditions for the convergence of the distributed iterative
algorithms based on the waterfilling solution and described in Section \ref%
{Section_Distributed Algorithms_WF}.

\section{Distributed Algorithms}

\label{Section_Distributed Algorithms for Nash Equilibria}In \cite[Theorem 2]%
{Scutari-Part I}, we proved that, under some (sufficient) conditions on
transmit powers, channels and network topology, the NE for game $%
{\mathscr{G}}%
$ is unique. Since there is no reason to expect a system to be initially at
the equilibrium, the concept of equilibrium has a useful meaning in practice
only if one is able to find a procedure that reaches such an equilibrium
from non-equilibrium states. In this section, we focus on algorithms that
converge to these equilibria.

Since we are interested in a decentralized implementation, where no
signaling among different users is allowed, we consider only totally
distributed iterative algorithms, where each user acts independently of the
others to optimize its own power allocation while perceiving the other users
as interference. The main issue of this approach is to guarantee the
convergence of such an iterative scheme. In the following, we propose two
alternative classes of totally distributed iterative algorithms along with
their convergence properties, namely: iterative algorithms based on the
waterfilling solution (\ref{WF_mask}), and iterative algorithms based on the
gradient projection mapping.

\subsection{Distributed algorithms based on waterfilling \label%
{Section_Distributed Algorithms_WF}}

So far, we have shown that the Nash equilibria of game $%
{\mathscr{G}}%
$\ are fixed-points (see (\ref{sym_WF-sistem})) of the waterfilling mapping
defined in (\ref{WF_mask}). Hence, to achieve these solutions by a
distributed scheme, it is natural to employ an iterative algorithm based on
the best response (\ref{WF_mask}). Based on this idea, we consider two
classes of iterative algorithms: \emph{sequential} algorithms, where the
users update their strategies sequentially according to a given schedule;
and \emph{simultaneous} algorithms, where all the users update their strategies
at the same time. In the following sections, we provide a formal description of
both algorithms and derive the conditions guaranteeing their convergence to
the unique NE of the game.

Before describing the proposed algorithms, we introduce the following
intermediate definitions. Given game ${%
{\mathscr{G}}%
,}$ let $\mathcal{D}_{q}\subseteq \{1,\cdots ,N\}$ denote the set $\{1,\ldots ,N\}$ (possibly)
deprived of the carrier indices that user $q$ would never use as the best
response set to any strategies adopted by the other users, for the given set of
transmit power and propagation channels \cite{Scutari-Part I}:
\begin{equation}
\mathcal{D}_{q}%
\triangleq%
\left\{ k\in \{1,\ldots ,N\}:\exists \text{ }\mathbf{p}_{-q}\in {%
{\mathscr{P}}%
}_{-q}\text{ such that }\left[ {\mathsf{WF}}_{q}\left( \mathbf{p}%
_{-q}\right) \right] _{k}\neq 0\right\} ,  \label{D_q}
\end{equation}%
with ${\mathsf{WF}}_{q}\left( \mathbf{\cdot }\right) $ defined in (\ref%
{WF_mask}) and ${%
{\mathscr{P}}%
}_{-q}%
\triangleq%
{%
{\mathscr{P}}%
}_{1}\times \cdots \times {%
{\mathscr{P}}%
}_{q-1}\times {%
{\mathscr{P}}%
}_{q+1}\times \cdots \times {%
{\mathscr{P}}%
}_{Q}$. In Part I of the paper \cite{Scutari-Part I}, we provided an
iterative procedure to estimate such a set. We also introduce the matrix $%
\mathbf{H}^{\max }\in
\mathbb{R}
^{Q\times Q}$, defined as
\begin{equation}
\left[ \mathbf{H}^{\max }\right] _{qr}%
\triangleq%
\left\{
\begin{array}{ll}
\Gamma _{q}\max\limits_{k\in \mathcal{D}_{q}\cap \mathcal{D}_{r}}\dfrac{|%
\bar{H}_{rq}(k)|^{2}}{|\bar{H}_{qq}(k)|^{2}}\dfrac{d_{qq}^{\gamma }}{%
d_{rq}^{\gamma }}\dfrac{P_{r}}{P_{q}}, & \text{if }\ r\neq q, \\
0, & \text{otherwise,}%
\end{array}%
\right.  \label{Hmax}
\end{equation}%
with the convention that the maximum in (\ref{Hmax}) is zero if $\ \mathcal{D%
}_{q}\cap \mathcal{D}_{r}$ is empty.

\subsubsection{Sequential iterative waterfilling algorithm revisited}

\label{Sec:IWFA}

The sequential IWFA is an instance of the Gauss-Seidel scheme \cite%
{Bertsekas Book-Parallel-Comp}: All users update their own strategies
sequentially, performing the waterfilling solution (\ref{WF_mask}). The
algorithm is described in Algorithm $1$.\bigskip

\begin{algo}{Sequential Iterative
Waterfilling Algorithm}SSet $\mathbf{p}_{q}^{(0)}=$ any
feasible power allocation, $\forall q\in
\Omega$; \newline\texttt{for} $n=0:\mathrm{Nit}$ \newline
\begin{equation}
\,\,\,\,\mathbf{p}_{q}^{(n+1)}=\left\{
\begin{array}
[c]{ll}%
\mathsf{WF}%
_{q}\left( \mathbf{p}_{-q}^{(n)}\right)  , & \text{if
}(n+1)\,\text{mod}\,Q=q,\\
\mathbf{p}_{q}^{(n)}, & \text{otherwise},
\end{array}
\right.  \hspace{1cm} \forall q\in\Omega; %
\end{equation}\label{IWFA_op}
\newline
\texttt{end}\label{IWFA_Algo}
\end{algo}
\bigskip

The convergence of the algorithm is guaranteed under the following
sufficient conditions.

\begin{theorem}
\label{Theo-IWFA} Assume that the following condition is satisfied:%
\begin{equation}
\rho \left( \mathbf{H}^{\max }\right) <1,  \tag{C1}  \label{SF-IWFA}
\end{equation}%
where $\mathbf{H}^{\max }$ is defined in (\ref{Hmax}) and $\rho \left(
\mathbf{H}^{\max }\right) $ denotes the spectral radius\footnote{%
The spectral radius $\rho \left( \mathbf{H}\right) $ of the matrix $\mathbf{H%
},$ is defined as $\rho \left( \mathbf{H}\right) =\max \left\{ \left\vert
\lambda \right\vert :\lambda \right. $ $\left. \in \mathrm{eig}\left(
\mathbf{H}\right) \right\} $, with $\mathrm{eig}\left( \mathbf{H}\right) $
denoting the set of eigenvalues of $\mathbf{H}$ \cite{Horn85}.} of the
matrix $\mathbf{H}^{\max }.$ Then, as $\mathrm{Nit}\rightarrow \mathrm{%
\infty }$, the sequential IWFA described in Algorithm \ref{IWFA_Algo}
converges linearly to the unique NE of game ${%
{\mathscr{G}}%
}$, for any set of initial conditions belonging to ${%
{\mathscr{P}}%
}$ and for any updating schedule.
\end{theorem}

\begin{proof}
See Appendix \ref{proof of Theorem Theo-IWFA}.
\end{proof}

\bigskip

\noindent \textbf{Remark 1 $-$ Global convergence and uniqueness of the NE. \ }%
Even though the optimization problem (\ref{Rate Game}) is nonlinear,
condition (\ref{SF-IWFA}) guarantees the \emph{global }convergence of the
sequential IWFA, irrespective of the specific users' updating order. Moreover, the global asymptotic stability of the NE
implies also the uniqueness of the equilibrium. Condition (\ref{SF-IWFA})
indeed coincides with the uniqueness condition given in \cite[Corollary 1]%
{Scutari-Part I}.

\bigskip

To give additional insight into the physical interpretation of sufficient
conditions for the convergence of the sequential IWFA, we provide the
following corollaries of Theorem \ref{Theo-IWFA}.

\begin{corollary}
\label{Corollary-C1-C2}A sufficient conditions for (\ref{SF-IWFA}) is given
by one of the two following set of conditions:%
\begin{equation}
\hspace{-0.1cm}\dfrac{\Gamma _{q}}{w_{q}}\text{ }\!\!\dsum\limits_{r=1,r\neq
q}\max\limits_{k\in \mathcal{D}_{r}\cap \mathcal{D}_{q}}\left\{ \dfrac{|\bar{%
H}_{rq}(k)|^{2}}{|\bar{H}_{qq}(k)|^{2}}\right\} \dfrac{d_{qq}^{\gamma }}{%
d_{rq}^{\gamma }}\dfrac{P_{r}}{P_{q}}w_{r}<1,\text{ }\forall q\in \Omega ,%
\hspace{-0.15cm}\vspace{-0.2cm}\medskip  \tag{C2}  \label{SF_for_C1_a}
\end{equation}%
\begin{equation}
\dfrac{1}{w_{r}}\!\!\text{ }\dsum\limits_{q=1,q\neq r}\Gamma
_{q}\max\limits_{k\in \mathcal{D}_{r}\cap \mathcal{D}_{q}}\left\{ \dfrac{|%
\bar{H}_{rq}(k)|^{2}}{|\bar{H}_{qq}(k)|^{2}}\right\} \dfrac{d_{qq}^{\gamma }%
}{d_{rq}^{\gamma }}\dfrac{P_{r}}{P_{q}}w_{q}<1,\text{ }\forall r\in \Omega ,
\tag{C3}  \label{SF_for_C1_b}
\end{equation}%
where $\mathbf{w}\triangleq \lbrack w_{1},\ldots ,w_{Q}]^{T}$ is any
positive vector.
\end{corollary}

\begin{corollary}
The best vector $\mathbf{w}$ in (\ref{SF_for_C1_a})-(\ref{SF_for_C1_b}) is
given by the solution to the following geometric programming problem\vspace{-0.2cm}
\begin{equation}
\begin{array}{ll}
\limfunc{minimize}\limits_{\mathbf{w},t} & t \\
\limfunc{subject}\,\limfunc{to} & \dsum\limits_{r=1,r\neq
q}G_{rq}t^{-1}w_{q}^{-1}w_{r}\leq 1,\quad \forall q, \\
& \mathbf{w}>\mathbf{0,}\text{ }t>0,%
\end{array}%
\vspace{-0.3cm}  \label{GP}
\end{equation}%
where $G_{rq}$ is defined as
\begin{equation}
G_{rq}\triangleq \Gamma _{q}\max\limits_{k\in \mathcal{D}_{r}\cap \mathcal{D}%
_{q}}\left\{ \dfrac{|\bar{H}_{rq}(k)|^{2}}{|\bar{H}_{qq}(k)|^{2}}\right\}
\dfrac{d_{qq}^{\gamma }}{d_{rq}^{\gamma }}\dfrac{P_{r}}{P_{q}}\frac{w_{r}}{%
w_{q}},
\end{equation}%
if (\ref{SF_for_C1_a}) is used, or as
\begin{equation}
G_{rq}\triangleq \Gamma _{r}\max\limits_{k\in \mathcal{D}_{r}\cap \mathcal{D}%
_{q}}\left\{ \dfrac{|\bar{H}_{qr}(k)|^{2}}{|\bar{H}_{rr}(k)|^{2}}\right\}
\dfrac{d_{rr}^{\gamma }}{d_{qr}^{\gamma }}\dfrac{P_{q}}{P_{r}}\frac{w_{q}}{%
w_{r}},
\end{equation}%
if (\ref{SF_for_C1_b}) is used.
\end{corollary}

\noindent

Note that, as by direct product of the proof of Theorem \ref{Theo-IWFA}, one
can always choose the full set $\mathcal{D}_{q}=\{1,\ldots ,N\}$ in (\ref%
{SF-IWFA}) and (\ref{SF_for_C1_a})-(\ref{SF_for_C1_b}). However, less
stringent conditions are obtained by removing the unnecessary carriers,
i.e., the carriers that, for the given power budget and interference levels,
are never going to be used.

\bigskip

\noindent \textbf{Remark 2 $-$ Physical interpretation of convergence
conditions.} As already shown in Part I of the paper \cite{Scutari-Part I}
for the uniqueness conditions of the NE, the convergence of sequential IWFA
is guaranteed if the interferers are sufficiently far apart from the
destinations. \ In fact, from (\ref{SF-IWFA}) or (\ref{SF_for_C1_a})-(\ref%
{SF_for_C1_b}), one infers that, for any given set of channel realizations
and power constraints, there exists a distance beyond which the sequential
IWFA is guaranteed to converge, corresponding to the maximum level of
interference that may be tolerated by each receiver (as quantified, e.g., in
(\ref{SF_for_C1_a})) or that may be generated by each transmitter (as
quantified, e.g., in (\ref{SF_for_C1_b})). Interestingly, the presence of
spectral mask constraints does not affect the convergence capability of the
algorithm. Moreover, convergence condition (\ref{SF-IWFA}) (or (\ref%
{SF_for_C1_a})-(\ref{SF_for_C1_b})) has the same desired properties as the
uniqueness conditions obtained in Part I\ of the paper: It is robust against
the worst normalized channels $|\bar{H}_{rq}(k)|^{2}/|\bar{H}_{qq}(k)|^{2},$
since the subchannels corresponding to the highest ratios $|\bar{H}%
_{rq}(k)|^{2}/|\bar{H}_{qq}(k)|^{2}$ (and, in particular, the subchannels
where $|\bar{H}_{qq}(k)|^{2}$ is vanishing) do not necessarily affect (\ref%
{SF-IWFA}) (or (\ref{SF_for_C1_a})-(\ref{SF_for_C1_b})), as their subcarrier
indices may not belong to the set $\mathcal{D}_{q}$. This strongly relaxes
the convergence conditions.

\bigskip

We can generalize the sequential IWFA given in Algorithm \ref{IWFA_Algo} by
introducing a memory in the updating process, as given in Algorithm \ref%
{IWFA_Algo_MEMO}. We call this new algorithm \emph{smoothed }sequential
IWFA. \bigskip

\begin{algo}{Smoothed Sequential Iterative
Waterfilling Algorithm}SSet $\mathbf{p}_{q}^{(0)}=$ any
feasible power allocation and $\alpha_q\in[0,\,1)$, $\forall q\in
\Omega$; \newline\texttt{for} $n=0:\mathrm{Nit}$ \newline
\begin{equation}
\,\mathbf{p}_{q}^{(n+1)}=\left\{
\begin{array}
[c]{ll}%
\alpha_q\mathbf{p}_{q}^{(n)}+(1-\alpha_q)\mathsf{WF}%
_{q}\left( \mathbf{p}_{-q}^{(n)}\right)  , & \text{if
}(n+1)\,\text{mod}\,Q=q,\\
\mathbf{p}_{q}^{(n)}, & \text{otherwise},
\end{array}
\right.  \hspace{0.2cm} \forall q\in\Omega; %
\end{equation}\label{IWFA_op}
\newline
\texttt{end}\label{IWFA_Algo_MEMO}
\end{algo}

\bigskip

Each factor $\alpha _{q}\in \lbrack 0,1)$ in Algorithm \ref{IWFA_Algo_MEMO}
can be interpreted as a forgetting factor: The larger is $\alpha _{q}$, the
longer is the memory of the algorithm. The choice of each $\alpha _{q}$
depends on the channel stationarity and on possible channel fluctuations or
estimation errors. If the channel is fixed or highly stationary and there
are channel estimation errors that induce zero mean fluctuations on the
allocated power, it is convenient to take each $\alpha _{q}$ close to $1$,
to smooth out the undesired fluctuations. Conversely, when the channel is
rapidly varying, it is better to take a small $\alpha _{q}$. Interestingly,
the choice of $\{\alpha _{q}\}_{q\in \Omega }$ does not affect the
convergence property of the algorithm, as proved in the following.

\begin{theorem}
\label{Theo-IWFA_memory} Assume that conditions of Theorem \ref{Theo-IWFA}
are satisfied. Then, as $\mathrm{Nit}\rightarrow \mathrm{\infty }$, the
smoothed sequential IWFA described in Algorithm \ref{IWFA_Algo_MEMO}
converges linearly to the unique NE of game ${%
{\mathscr{G}}%
,}$ for any set of initial conditions in ${%
{\mathscr{P}}%
}$, updating schedule, and $\{\alpha _{q}\}_{q\in \Omega },$ with $\alpha
_{q}\in \left[ 0,1\right) $, $\forall q\in \Omega .$
\end{theorem}

\begin{proof}
See Appendix \ref{proof of Theorem Theo-IWFA}.
\end{proof}

\bigskip

\noindent \textbf{Remark 3 $-$ Comparison with previous results. }The
sequential IWFA described in Algorithm \ref{IWFA_Algo} generalizes the
well-known sequential iterative waterfilling algorithm originally proposed
by Yu et al. in \cite{Yu} and then studied in \cite{ChungISIT03}-\cite{Tse},
to the case in which the users have (possibly) different power budgets and
there are spectral mask constraints. In fact, the algorithm in \cite{Yu} can
be obtained as a special case of Algorithm \ref{IWFA_Algo}, by removing the
spectral mask constraints in each set ${\mathscr{P}}_{q}$ in (\ref%
{admissible strategy set}), (i.e., setting $p_{q}^{\max }(k)=+\infty ,$ $%
\forall k,q$) and replacing the waterfilling operator in (\ref{WF_mask})
with the classical waterfilling solution%
\begin{equation}
\begin{array}{c}
\mathsf{WF}%
_{q}\left( \mathbf{p}_{1},\ldots ,\mathbf{p}_{q-1},\mathbf{p}_{q+1},\ldots ,%
\mathbf{p}_{Q}\right)
\triangleq%
\left( \mu _{q}\mathbf{1}-%
\boldsymbol{\mathsf{insr}}%
_{q}(\mathbf{p}_{-q})\right) ^{+}%
\end{array}%
,\quad  \label{WF-solution}
\end{equation}%
where $\left( x\right) ^{+}=\max (0,x)$ and $%
\mathsf{insr}%
_{q}$ is defined in (\ref{INR_q}).

The convergence of the sequential IWFA based on the mapping (\ref%
{WF-solution}) has been studied in a number of works, each time obtaining
milder convergence conditions. Specifically, in \cite{Yu}, the authors
provided sufficient conditions for the existence of a NE and the convergence
of the sequential IWFA, for a game composed by two players. This was later
generalized to an arbitrary number of players in \cite{ChungISIT03}-\cite%
{Scutari-Barbarossa-SPAWC03}. In \cite{Tse}, the case of flat-fading
channels was considered. Interestingly, although derived under stronger
constraints, incorporating for example spectral mask constraints, our
convergence conditions have a broader validity than those obtained in \cite%
{Yu}-\cite{Tse}, as shown in the following.\footnote{%
We summarize the main results of \cite{Yu}-\cite{Scutari-Barbarossa-SPAWC03}
using our notation.}

\begin{corollary}
\label{Corollary:Conditions of the others}Sufficient conditions for (\ref%
{SF_for_C1_a}) are \cite{Yu, ChungISIT03, Scutari-Barbarossa-SPAWC03}%
\begin{equation}
\Gamma _{q}\max\limits_{k\in \{1,\ldots ,N\}}\left\{ \dfrac{|\bar{H}%
_{rq}(k)|^{2}}{|\bar{H}_{qq}(k)|^{2}}\right\} \dfrac{d_{qq}^{\alpha }}{%
d_{rq}^{\alpha }}\dfrac{P_{r}}{P_{q}}<\dfrac{1}{Q-1},\hspace{1.5cm}\forall
\text{ }r,q\neq r\in \Omega ,  \tag{C4}  \label{Chung SF_}
\end{equation}%
or \cite{Yamashitay-Luo}%
\begin{equation}
\Gamma _{q}\max\limits_{k\in \{1,\ldots ,N\}}\left\{ \dfrac{|\bar{H}%
_{rq}(k)|^{2}}{|\bar{H}_{qq}(k)|^{2}}\right\} \dfrac{d_{qq}^{\alpha }}{%
d_{rq}^{\alpha }}\dfrac{P_{r}}{P_{q}}<\dfrac{1}{2Q-3},\hspace{1.5cm}\forall
\text{ }r,q\neq r\in \Omega .  \tag{C5}  \label{Luo-Yamashitay}
\end{equation}

In the case of flat-fading channels (i.e., $\bar{H}_{rq}(k)=\bar{H}_{rq}$, $%
\forall r,q$), condition (\ref{SF_for_C1_a}) becomes \cite{Tse}%
\begin{equation}
\Gamma _{q}\sum_{r=1,r\neq q}^{Q}\dfrac{|\bar{H}_{rq}|^{2}}{|\bar{H}%
_{qq}|^{2}}\dfrac{d_{qq}^{\alpha }}{d_{rq}^{\alpha }}\dfrac{P_{r}}{P_{q}}%
<1,\qquad \forall q\in \Omega .  \label{Tse-cond}
\end{equation}
\end{corollary}

Recently, alternative sufficient conditions for the convergence of
sequential IWFA as given in Algorithm \ref{IWFA_Algo} were independently
given in \cite{Luo-Pang}.\footnote{%
We thank Prof. Facchinei, who kindly brought to our attention reference \cite%
{Luo-Pang}, after this paper was completed.} Specifically, the sequential
IWFA was proved to converge to the unique NE of the game if the following
condition is satisfied:\footnote{%
We write conditions of \cite{Luo-Pang} using our notation.}
\begin{equation}
\rho \left( \mathbf{\Upsilon }\right) <1,  \tag{C6}  \label{Luo}
\end{equation}%
where $\rho \left( \mathbf{\Upsilon }\right) $ denotes the spectral radius
of the matrix $\mathbf{\Upsilon }\triangleq \left( \mathbf{I}-\mathbf{H}_{%
\text{low}}^{\max }\right) ^{-1}\mathbf{H}_{\text{upp}}^{\max }$, with $%
\mathbf{H}_{\text{low}}^{\max }$ and $\mathbf{H}_{\text{upp}}^{\max }$
denoting the strictly lower and strictly upper triangular part of the matrix
$\mathbf{H}^{\max }$, respectively, with $\mathbf{H}^{\max }$ defined, in
our notation, as in (\ref{Hmax}), where each $\mathcal{D}_{q}$ is replaced
by the full set $\{1,\ldots ,N\}.$

\begin{figure}[tbp]
\vspace{-0.8cm}
\par
\begin{center}
\subfigure[] {\includegraphics[height=7cm, width=8.9cm]{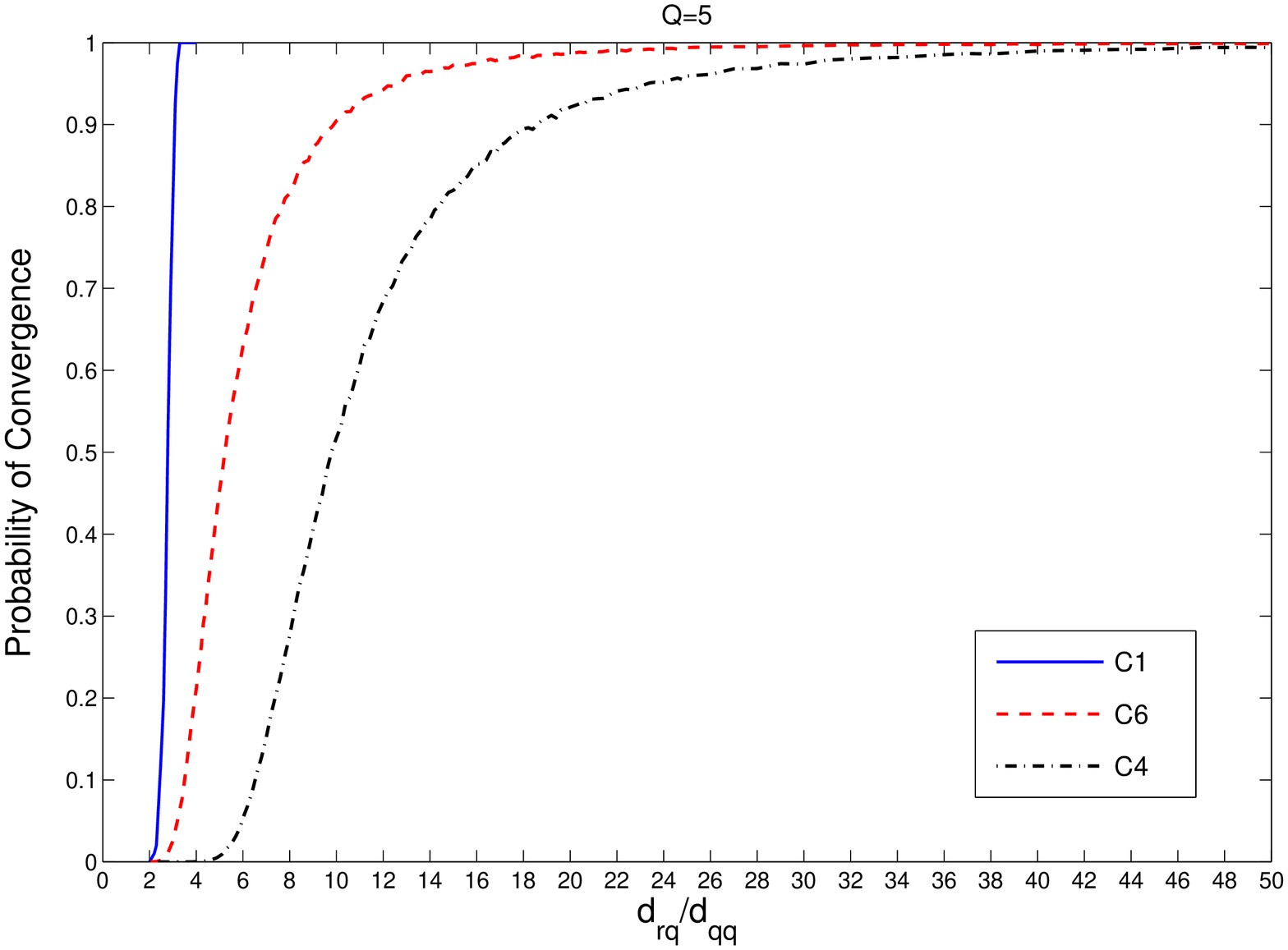}}
\hspace{-1cm}%
\subfigure[] {
\includegraphics[height=7cm,width=8.9cm]{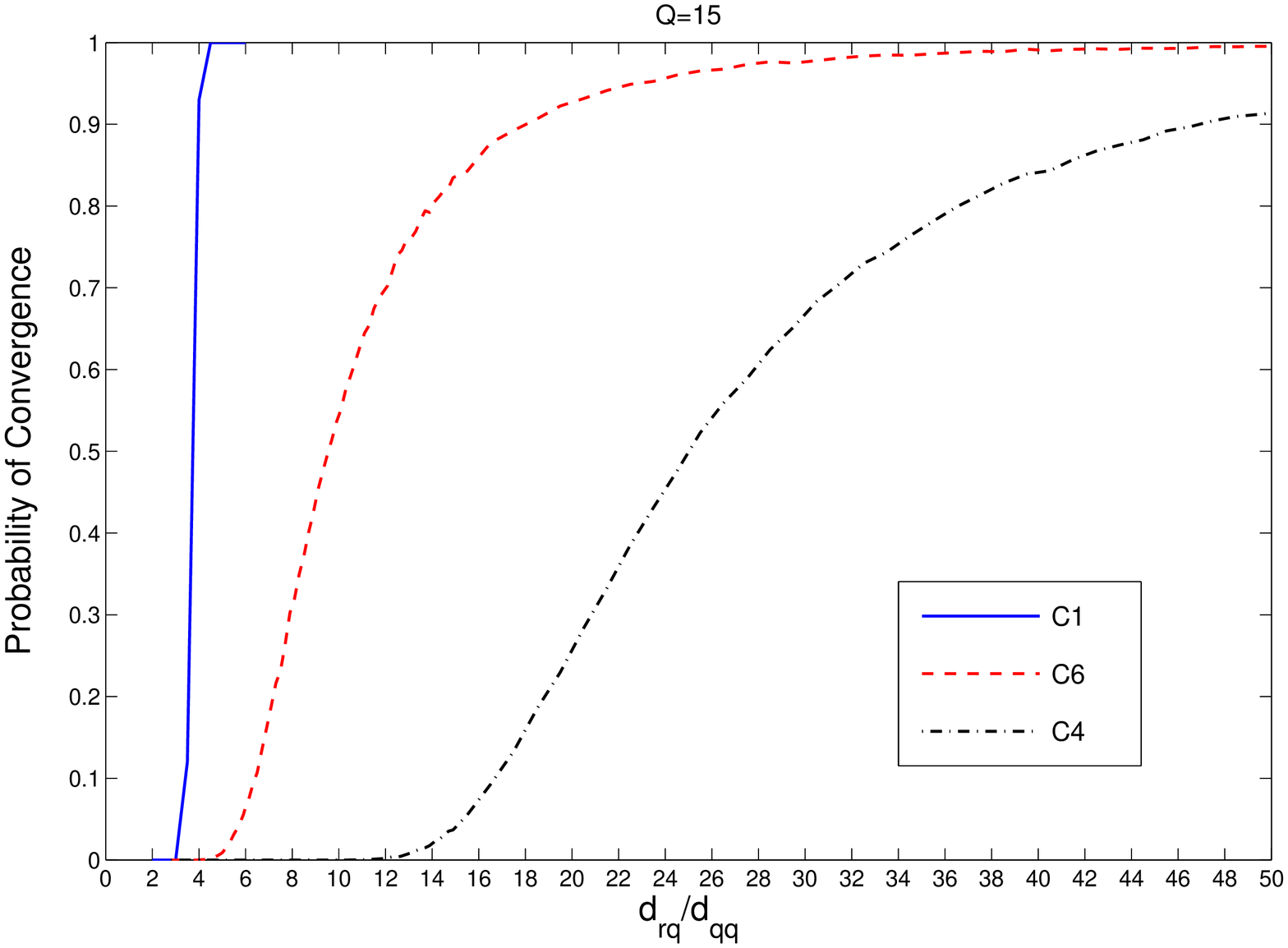}}\vspace{-1cm}
\end{center}
\caption{{\protect\scriptsize Probability of (\protect\ref{SF-IWFA}), (%
\protect\ref{Chung SF_}) and (\protect\ref{Luo}) versus $d_{rq}/d_{qq}$; $%
Q=5 $ [subplot (a)], $Q=15$ [subplot (b)], $\protect\gamma =2.5,$ $d_{rq}$ $%
=d_{qr},$ $d_{rr}=d_{qq}=1$, $P_q=P_r$, $\Gamma_{q} =1$, $P_q/\protect\sigma%
_q^2=7$dB, $P_r/(\protect\sigma_q^2 d_{rq}^\protect\gamma)=3$dB, $\forall
r,q\in \Omega,$ $\boldsymbol{w}=1$.}}
\label{fig:Comparison_SF_Conds}
\end{figure}

As an example, in Figure \ref{fig:Comparison_SF_Conds}, we compare the range
of validity of our convergence condition (\ref{SF-IWFA}) with that of (\ref%
{Chung SF_}) and (\ref{Luo}), over a set of channel impulse responses
generated as vectors composed of i.i.d. complex Gaussian random variables
with zero mean and unit variance. In the figure, we plot the probability
that conditions (\ref{SF-IWFA}), (\ref{Chung SF_}) and (\ref{Luo}) are
satisfied versus the ratio $d_{rq}/d_{qq},$ which measures how far apart are
the interferers from the destination, with respect to the intended source.
In Figure \ref{fig:Comparison_SF_Conds}(a) we consider a system composed by $%
Q=5$ users, and in Figure \ref{fig:Comparison_SF_Conds}(b) a system with
$Q=15$ links. For the sake of simplicity, to limit the number of free
parameters, we assumed $d_{rq}=d_{qr},$ $P_{q}=P_{r}$ $\forall r,q,$ and $%
\mathbf{w}=\mathbf{1}.$ We tested our condition considering the set $%
\mathcal{D}_{q},$ obtained using the algorithm given in \cite{Scutari-Part I}%
. We can see, from Figure \ref{fig:Comparison_SF_Conds}, that the
probability of guaranteeing convergence increases as the distance of the
interferers, normalized to the source-destination distance, increases (i.e.,
the ratio $d_{rq}/d_{qq}$ increases). Interestingly, the probability that (%
\ref{SF-IWFA}) is satisfied, differently from (\ref{Chung SF_}) and (\ref%
{Luo}), exhibits a neat threshold behavior as it transits very rapidly from
the non-convergence guarantee to the almost certain convergence, as the
ratio $d_{rq}/d_{qq}$ increases by a small percentage. This shows that the
convergence conditions depend, fundamentally, on the interferers distance,
rather than on the channel realizations. Finally, it is worthwhile noticing
that our conditions have a broader validity than (\ref{Chung SF_}) and (\ref%
{Luo}). As an example, for a system with probability of guaranteeing
convergence of $0.99$ and $Q=15,$ conditions (\ref{SF-IWFA}) only require $%
d_{rq}/d_{qq}\simeq 4.2,$ whereas conditions (\ref{Chung SF_}) and (\ref{Luo}%
) require $d_{rq}/d_{qq}>50$ and $d_{rq}/d_{qq}\simeq 40,$ respectively.
Furthermore, comparing Figure \ref{fig:Comparison_SF_Conds}(a) with Figure %
\ref{fig:Comparison_SF_Conds}(b), one can see that this difference increases
as the number $Q$ of links increases.

\bigskip

\noindent \textbf{Remark 4 $-$ Distributed nature of the algorithm.} The
sequential IWFA as described in Algorithms \ref{IWFA_Algo} and \ref%
{IWFA_Algo_MEMO} can be implemented in a distributed way, since each user,
to maximize its own rate, needs only to measure the PSD of the thermal noise
plus the overall MUI (see (\ref{SINR})). However,
despite its appealing properties, the algorithm may suffer from slow
convergence if the number of users in the network is large, as we will also
show in Section \ref{Sec:Convergence_rate}. This drawback is due to the
sequential schedule in the users' updates, wherein each user, to choose its
own strategy, is forced to wait for all the other users scheduled before.
Moreover, although distributed, both algorithms require that all users share
a prescribed updating schedule. This requires a centralized synchronization
mechanism that determines the order and the update times of the users. We
show next how to remove these limitations.

\subsubsection{Simultaneous iterative waterfilling algorithm \label%
{Section_Simultaneous Distributed Algorithms}}

To overcome the main limitation of sequential IWFAs given in Algorithms \ref%
{IWFA_Algo} and \ref{IWFA_Algo_MEMO}, we consider in this section the
simultaneous version of the IWFA, called \emph{simultaneous }IWFA. The
algorithm is an instance of the Jacobi scheme \cite{Bertsekas
Book-Parallel-Comp}: At each iteration, all users update their own power
allocation \emph{simultaneously}, performing the waterfilling solution (\ref%
{WF_mask}), given the interference generated by the other users in the \emph{%
previous} iteration. Stated in mathematical terms, the proposed algorithm is
described in Algorithm \ref{SIWFA_algo} \cite{Scutari_Thesis, Scutari_ISIT06}.\bigskip

\begin{algo}{\hspace{0.6cm} Simultaneous Iterative Waterfilling Algorithm }
SSet $\mathbf{p}_{q}^{(0)}=$ any feasible power allocation, $\forall q\in
\Omega;$\vspace{0.1cm}\newline
\texttt{for} $n=0:\mathrm{Nit}$
\begin{equation}
\,\,\,\,\mathbf{p}_{q}^{(n+1)}=
{\mathsf{{WF}}}%
_{q}\left( \mathbf{p}_{1}^{(n)},\ldots ,\mathbf{p}_{q-1}^{(n)},\mathbf{p}%
_{q+1}^{(n)},\ldots ,\mathbf{p}_{Q}^{(n)}\right),\hspace{1cm}\forall
q\in\Omega, \label{SIWFA_op}
\end{equation}%
\texttt{end}
\label{SIWFA_algo}
\end{algo}

\bigskip

As for the sequential IWFA, also in the simultaneous IWFA we can introduce a
memory in the updating process and obtain the so-called \emph{smoothed}
simultaneous IWFA, as described in Algorithm \ref{SIWFA_algo_MEMO} \cite%
{Scutari_Thesis, Scutari_ISIT06}.\bigskip

\begin{algo}{\hspace{0.6cm} Smoothed Simultaneous Iterative Waterfilling Algorithm }
SSet $\mathbf{p}_{q}^{(0)}=$ any feasible power allocation and $\alpha_q\in[0,\,1)$, $\forall q\in
\Omega$;\vspace{0.1cm}\newline
\texttt{for} $n=0:\mathrm{Nit}$
\begin{equation}
\,\,\,\,\mathbf{p}_{q}^{(n+1)}=\alpha_q \mathbf{p}_{q}^{(n)}+(1-\alpha_q)
{\mathsf{{WF}}}_{q}\left( \mathbf{p}_{1}^{(n)},\ldots ,\mathbf{p}_{q-1}^{(n)},\mathbf{p}%
_{q+1}^{(n)},\ldots ,\mathbf{p}_{Q}^{(n)}\right),\hspace{1cm}\forall
q\in\Omega, \label{SIWFA_op}
\end{equation}%
\texttt{end}
\label{SIWFA_algo_MEMO}
\end{algo}

\bigskip

Interestingly, both Algorithm \ref{SIWFA_algo} and \ref%
{SIWFA_algo_MEMO} are guaranteed to globally converge to the unique NE of
the game, under the same sufficient conditions of the sequential IWFA, as
proved in the following.

\begin{theorem}
\label{Theo-SIWFA} Assume that conditions of Theorem \ref{Theo-IWFA} are
satisfied. Then, as $\mathrm{Nit}\rightarrow \mathrm{\infty }$, \ the
simultaneous IWFAs described in Algorithm \ref{SIWFA_algo} and Algorithm \ref%
{SIWFA_algo_MEMO} converge linearly to the unique NE of game ${%
{\mathscr{G}}%
,}$ for any set of initial conditions in ${%
{\mathscr{P}}%
}$ and $\{\alpha _{q}\}_{q\in \Omega },$ with $\alpha _{q}\in \left[
0,1\right) $, $\forall q\in \Omega $.
\end{theorem}

\begin{proof}
See Appendix \ref{proof of Theorem Theo-SIWFA}.
\end{proof}

\bigskip

\noindent Additional (weaker) convergence conditions for Algorithm \ref%
{SIWFA_algo} and \ref{SIWFA_algo_MEMO} are given next. Introducing
the matrix $\mathbf{H}(k)\in
\mathbb{R}
^{Q\times Q},$ defined as
\begin{equation}
\lbrack \mathbf{H}(k)]_{qr}%
\triangleq%
\left\{
\begin{array}{ll}
\Gamma _{q}\dfrac{|\bar{H}_{rq}(k)|^{2}}{|\bar{H}_{qq}(k)|^{2}}\dfrac{%
d_{qq}^{\gamma }}{d_{rq}^{\gamma }}\dfrac{P_{r}}{P_{q}},\quad & \text{if \ }%
k\in \mathcal{D}_{q}\cap \mathcal{D}_{r},\text{ and }q\neq r, \\
\text{ }0, & \text{otherwise,}%
\end{array}%
\right.  \label{H(k)_qr}
\end{equation}%
we have the following theorem.

\begin{theorem}
\label{Theorem Theo-SIWFA_general}Assume that the following conditions are
satisfied:%
\begin{equation}
\rho ^{1/2}\left( \mathbf{H}^{T}(k)\mathbf{H}(k)\right) <1,\text{\qquad }%
\forall k\in \{1,\ldots ,N\},  \label{C4}
\end{equation}%
where $\mathbf{H}(k)$ is defined in (\ref{H(k)_qr}). Then, as $\mathrm{Nit}%
\rightarrow \mathrm{\infty }$, the sequential IWFA\footnote{%
Condition (\ref{C4}) is sufficient also for the convergence of the smoothed
simultaneous IWFA described in Algorithm \ref{SIWFA_algo_MEMO}, provided
that the second hand-side of (\ref{C4}) is replaced by $\epsilon =\frac{%
1-\max_{q\in \Omega }\alpha _{q}}{1-\min_{q\in \Omega }\alpha _{q}}\leq 1.$}
described in Algorithm \ref{SIWFA_algo} converges linearly to the unique NE
of game ${%
{\mathscr{G}}%
,}$ for any set of initial conditions in ${%
{\mathscr{P}}%
}.$
\end{theorem}

\begin{proof}
See Appendix \ref{proof of Theorem Theo-SIWFA_general}.
\end{proof}

\bigskip

\noindent \textbf{Remark 5 $-$ Sequential versus simultaneous IWFA.} Since both simultaneous IWFAs in Algorithms \ref%
{SIWFA_algo} and \ref{SIWFA_algo_MEMO} are still based on the waterfilling
solution (\ref{WF_mask}), they keep the most appealing features of the
sequential IWFA, namely its low-complexity and distributed nature. In fact,
as in the sequential IWFA, also in the simultaneous IWFA each user only
needs to locally measure the PSD of the interference received from the other
users and water-pour over this level. In addition, thanks to the
Jacobi-based update, all the users are allowed to choose their optimal power
allocation \textit{simultaneously}. Hence, the simultaneous IWFA is expected
to be faster than the sequential IWFA, especially if the number of active
users in the network is large. We formalize this intuition in the next
section. 

\subsubsection{Asymptotic convergence rate}

\label{Sec:Convergence_rate}

In this section we provide an upper bound of the convergence rate of both
(smoothed) sequential and simultaneous IWFAs. The convergence rate can be
either measured on the average or for the worst possible initial vector $%
\mathbf{p}^{(0)}.$ In the following we focus on the latter approach,
introducing the asymptotic convergence exponent.

Denoting by $\mathbf{p}^{\star }$ and $\mathbf{p}^{(n)}$ the NE of game ${%
{\mathscr{G}}%
}$ and the power allocation vector obtained by the proposed algorithm at the
$n$-th iteration, respectively, the distance between $\mathbf{p}^{(n)}$ and $%
\mathbf{p}^{\star }$ can be measured by some vector norm $\left\Vert \mathbf{%
p}^{(n)}-\mathbf{p}^{\star }\right\Vert ,$ which is to be compared with the
initial distance $\left\Vert \mathbf{p}^{(0)}-\mathbf{p}^{\star }\right\Vert
.$ This leads to the following asymptotic convergence exponent for the
worst-case convergence rate \cite{Ortega}:
\begin{equation}
d=-\sup_{\mathbf{p}^{(0)}\neq \mathbf{p}^{\star }}\lim_{n\rightarrow \infty }%
\frac{1}{n}\ln \left( \dfrac{\left\Vert \mathbf{p}^{(n)}-\mathbf{p}^{\star
}\right\Vert }{\left\Vert \mathbf{p}^{(0)}-\mathbf{p}^{\star }\right\Vert }%
\right) .  \label{conv_factor}
\end{equation}%
Since for large $n$
\begin{equation}
\left\Vert \mathbf{p}^{(n)}-\mathbf{p}^{\star }\right\Vert \simeq Ce^{-dn},
\end{equation}%
where $C$ is a constant that depends on the initial conditions, $d$ gives
the (asymptotic) number of iterations for the error to decrease by the
factor $1/e$ (for the worst possible initial vector).

Since the waterfilling operator is not a monotone mapping, only (upper)
bounds for the asymptotic convergence exponent can be obtained \cite%
{Bertsekas-Tsitsklis-convergence-mapping, Tsitsiklis-Jacobi-GaussSeidel}, as
given in the following.

\begin{proposition}
\label{convergence rate}Let $d_{\text{seq}}^{\text{\,low}}$ and $d_{\text{sim}%
}^{\text{\,low}}$ be lower bound of $d$ in (\ref{conv_factor}) obtained using
(smoothed) sequential IWFA in Algorithm \ref{IWFA_Algo_MEMO} and (smoothed)
simultaneous IWFA in Algorithm \ref{SIWFA_algo_MEMO}, respectively. Under
condition (\ref{SF_for_C1_a}) of Corollary \ref{Corollary-C1-C2}, we have
\begin{align}
d_{\text{seq}}^{\text{\,low}}& =-\log \left( \max_{q\in \Omega }\left\{ \alpha
_{q}+(1-\alpha _{q})\dfrac{\Gamma _{q}}{w_{q}}\dsum\limits_{r\neq
q}\max\limits_{k\in \mathcal{D}_{r}\cap \mathcal{D}_{q}}\left\{ \dfrac{|\bar{%
H}_{rq}(k)|^{2}}{|\bar{H}_{qq}(k)|^{2}}\right\} \dfrac{d_{qq}^{\gamma }}{%
d_{rq}^{\gamma }}\dfrac{P_{r}}{P_{q}}w_{r}\right\} \right) ,
\label{conv_rate_comparison_1} \\
d_{\text{sim}}^{\text{\,low}}& =Q\text{ }d_{\text{seq}}^{\text{\,low}}.
\label{conv_rate_comparison_2}
\end{align}
\end{proposition}

\begin{proof}
The proof follows directly from Proposition 2 in Appendix \ref%
{Appendix_WF_project}.
\end{proof}

\bigskip

\noindent \textbf{Remark 6 $-$ Convergence speed}. Expression (\ref{conv_rate_comparison_1}) shows
that the convergence speed of the algorithms depends, as expected, on the
memory factors $\{\alpha _{q}\}_{q\in \Omega }$ and on the level of
interference. Given $\{\alpha _{q}\}_{q\in \Omega }$, the convergence speed
increases as the interference level decreases.
Since $d_{\text{sim}}^{\text{\,low}}$ and $d_{\text{seq}}^{\text{\,low}}$ are
only bounds of the asymptotic convergence exponent, a comparison between the
sequential IWFA and the simultaneous IWFA by $d_{\text{sim}}^{\text{\,low}}$
and $d_{\text{seq}}^{\text{\,low}}$ might not be fair. These bound becomes
meaningful if $d_{\text{sim}}^{\text{\,low}}$ and $d_{\text{seq}}^{\text{\,low}}$
approximate with equality $d_{\text{sim}}$ and $d_{\text{seq}},$
respectively, for some initial conditions (cf. \cite%
{Bertsekas-Tsitsklis-convergence-mapping}).

\begin{figure}[tbph]
\vspace{-0.5cm}
\par
\begin{center}
\includegraphics[trim=0.000000in 0.000000in 0.000000in
-0.212435in, height=8cm]{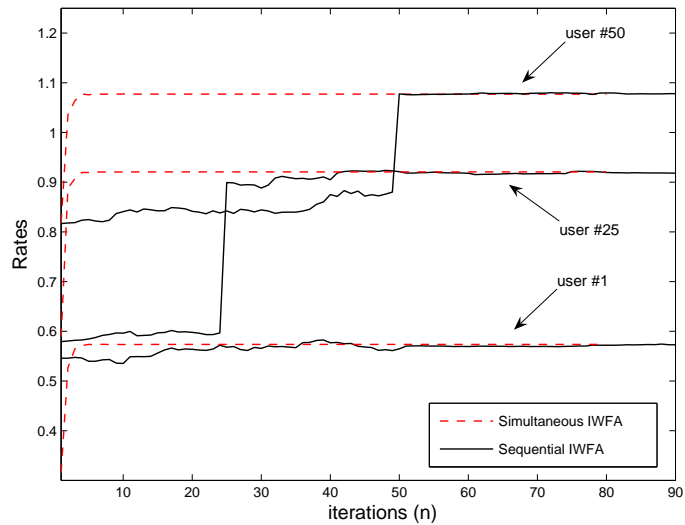}
\end{center}
\par
\vspace{-0.8cm}
\caption{{\protect\scriptsize Rates of the users versus iterations:
sequential IWFA (solid line curves), simultaneous IWFA (dashed line curves),
$Q=50$, $\protect\gamma =2.5,$ $d_{rq}$ $=d_{qr},$ $d_{rr}=d_{qq}=1$, $%
P_q=P_r$, $\Gamma_{q} =1$, $P_q/\protect\sigma_q^2=7$dB, $P_r/(\protect\sigma%
_q^2 d_{rq}^\protect\gamma)=3$dB, $\forall r,q\in \Omega$.}}
\label{SIWFA-IWFA}
\end{figure}
In Figure \ref{SIWFA-IWFA} we compare the performance of the sequential and
simultaneous IWFA, in terms of convergence speed. We consider a network
composed of 50 links and we show the rate evolution of three of the links
corresponding to the sequential IWFA and simultaneous IWFA as a function of
the iteration index $n$ as defined in Algorithms \ref{IWFA_Algo} and \ref%
{SIWFA_algo}. To make the figure not excessively overcrowded, we report only
the curves of $3$ out of $50$ links. As expected, the sequential IWFA is
slower than the simultaneous IWFA, especially if the number of active links $%
Q$ is large, since each user is forced to wait for all the users
scheduled in advance, before updating its own power allocation.

\subsection{Distributed algorithms based on gradient projection \label%
{Sec:Simultaneous Gradient Projection based Algorithm}}

In this section we propose two alternative distributed algorithms based on
the Gradient Projection mapping. The first algorithm is an instance of the
Jacobi scheme, whereas the second one is based on the Gauss-Seidel
procedure. Both algorithms come out from an interesting interpretation of the
Nash equilibria in (\ref{sym_WF-sistem}) as solutions to a proper Nonlinear
Variational Inequality (NVI) problem \cite[Sec. 1.4.2]{Facchinei}, as we
show next. The NVI problem is defined as follows. Given a subset $\mathcal{X}%
\subset
\mathbb{R}
^{n}$ of the Euclidean $n$-th dimensional space $%
\mathbb{R}
^{n}$ and a mapping $\mathbf{f}:%
\mathbb{R}
^{n}\mapsto
\mathbb{R}
^{n},$ the (nonlinear) \emph{variational inequality} is to find a vector $%
\mathbf{x}^{\star }\in \mathcal{X}$ such that \cite[Def. 1.1.1]{Facchinei}%
\vspace{-0.2cm}
\begin{equation}
\text{(NVI)\hspace{3cm}}\left( \mathbf{x-x}^{\star }\right) ^{T}\mathbf{f}(%
\mathbf{x}^{\star })\geq 0,\quad \quad \forall \mathbf{x}\in \mathcal{X}%
\text{.\hspace{3cm}}\vspace{-0.2cm}  \label{VI}
\end{equation}

All the Nash equilibria in (\ref{sym_WF-sistem}) of game ${%
{\mathscr{G}}%
}$, can be written as solutions to a NVI problem. In fact, a feasible
strategy profile $\mathbf{p}^{\star }=[\mathbf{p}_{1}^{\star T},\ldots ,%
\mathbf{p}_{Q}^{\star T}]^{T}$ satisfies (\ref{sym_WF-sistem}) if and only
if the following necessary and sufficient optimality conditions hold true
\cite{Boyd}:\footnote{%
Observe that, given the strategy profiles of the other players, the optimal
strategy of each player is a solution to the convex optimization problem
defined in (\ref{Rate Game}), whose optimality conditions, for any given $q$
and $\mathbf{p}_{-q}^{{\star }},$ can be written as in (\ref{KKT-NE}) \cite[%
Sec. 4.2.3]{Boyd}.}
\begin{equation}
(\mathbf{p}_{q}\mathbf{-p}_{q}^{\star })^{T}(-%
\nabla%
_{q}R_{q}(\mathbf{p}_{q}^{\star },\mathbf{p}_{-q}^{\star }))\geq 0,\quad
\forall \mathbf{p}_{q}\in {\mathscr{P}}_{q}\text{, \ }\forall q\in \Omega ,
\label{KKT-NE}
\end{equation}%
where $%
\nabla%
_{q}R_{q}(\mathbf{p}_{q}^{\star },\mathbf{p}_{-q}^{\star })$ denotes the
gradient vector of $R_{q}$ with respect to $\mathbf{p}_{q},$ evaluated in $(%
\mathbf{p}_{q}^{\star },\mathbf{p}_{-q}^{\star }),$ and ${\mathscr{P}}_{q}$
is defined in (\ref{admissible strategy set}). Comparing (\ref{VI}) with (%
\ref{KKT-NE}), it is straightforward to see that a strategy profile $\mathbf{%
p}^{\star }$ is a NE of ${%
{\mathscr{G}}%
}$ if and only if it is a solution to the NVI problem (\ref{VI}), with the
following identifications:
\begin{eqnarray}
&&{\mathcal{X}}={\mathcal{X}}_{1}\times \ldots \times \mathcal{X}_{Q}\,\,{%
\longleftrightarrow }\,\,{\mathscr{P}=\mathscr{P}}_{1}\times \ldots \times {%
\mathscr{P}}_{Q}, \\
&&\mathbf{x}=\mathbf{[x}_{1}^{T},\ldots ,\mathbf{x}_{Q}^{T}]^{T}\,\,{%
\longleftrightarrow }\,\,\mathbf{p}=[\mathbf{p}_{1}^{T},\ldots ,\mathbf{p}%
_{Q}^{T}]^{T} \\
&&f_{q}(\mathbf{x}_{q},\mathbf{x}_{-q})\,\,{\longleftrightarrow }\,-\,%
\nabla%
_{q}R_{q}(\mathbf{p}_{q},\mathbf{p}_{-q}),
\end{eqnarray}%
%
%
%
%
%
%
%
%
%
%
%
%
%
%
%
%
%
%
where $f_{q}(\mathbf{x})$ denotes the $q$-th component of $\mathbf{f(x)}%
=[f_{1}(\mathbf{x}),\ldots ,f_{Q}(\mathbf{x})]^{T}.\mathbf{\ }$In fact, if (%
\ref{KKT-NE}) is satisfied for each $q$, then summing over $q$, (\ref{VI})
follows. Conversely, assume that (\ref{VI}) holds true for some $\mathbf{p}%
^{\star }$. Then, for any given $q,$ choosing $\mathbf{p}_{-q}=\mathbf{p}%
_{-q}^{\star }$ and $\mathbf{p}_{q}\in {\mathscr{P}}_{q},$ we have $\mathbf{p%
}%
\triangleq%
(\mathbf{p}_{q},\mathbf{p}_{-q}^{\star })\in {\mathscr{P}}$ and $\left(
\mathbf{p-p}^{\star }\right) ^{T}[-%
\nabla%
_{1}^{T}R_{1}(\mathbf{p}^{\star }),\ldots ,-%
\nabla%
_{Q}^{T}R_{Q}(\mathbf{p}^{\star })]^{T}=$ $(\mathbf{p}_{q}\mathbf{-p}%
_{q}^{\star })^{T}(-%
\nabla%
_{q}R_{q}(\mathbf{p}_{q}^{\star },\mathbf{p}_{-q}^{\star }))\geq 0,$ $%
\forall \mathbf{p}_{q}\in {\mathscr{P}}_{q}.$ 

Building on the equivalence between (\ref{VI}) and (\ref{KKT-NE}) \cite%
{Facchinei, Nagurney}, we can obtain distributed algorithms that reach the
Nash equilibria of game ${%
{\mathscr{G}}%
}$ by looking for algorithms that solve the NVI problem in (\ref{VI}). A
similar approach was already followed in \cite{Yamashitay-Luo}, where the
equivalence between the Nash equilibria of a DSL game that is a special case
of ${%
{\mathscr{G}}%
}$ and the solutions to a proper nonlinear complementary problem \cite%
{Facchinei} was shown. However, the algorithms proposed in \cite%
{Yamashitay-Luo} to compute the NE solutions, in general, cannot be
implemented in a distributed way, since they need a centralized control
having access to all channel state information and to the PSD of all users.
Differently from \cite{Yamashitay-Luo}, we exploit the equivalence between (%
\ref{VI}) and (\ref{KKT-NE}) and propose two alternative totally distributed
algorithms that do not require any centralized control to be implemented and
have the same computational complexity as the IWFAs. To this end, we need
the following intermediate result that comes directly from the NVI
formulation in (\ref{KKT-NE}) \cite[Prop. 5.1]{Bertsekas Book-Parallel-Comp}.

\begin{lemma}
\label{Lemma:GP_Mapping}Let $\beta $ be a positive scalar and $\{\mathbf{G}%
_{q}\}_{q\in \Omega }$ be a set of symmetric positive definite matrices. A\
vector $\mathbf{p}^{\star }=[\mathbf{p}_{1}^{\star T},\ldots ,\mathbf{p}%
_{Q}^{\star T}]^{T}$ is a NE of game ${%
{\mathscr{G}}%
}$ if and only if it is a fixed point of the following mapping:\footnote{%
The mapping in (\ref{GP_mapping}) always admits at least one fixed point,
since it satisfies Brouwer's fixed point theorem \cite[Theorem 4.2.5]%
{Brouwer}. In fact, each set ${\mathscr{P}}_{q}$ is compact and convex, and
the mapping $\mathbf{T}_{\mathbf{G}}(\mathbf{p})=(\mathbf{T}_{\mathbf{G}%
_{1}}(\mathbf{p}))_{q\in \Omega }$ in (\ref{GP_mapping}) is continuous on ${%
\mathscr{P}}={\mathscr{P}}_{1}\times \cdots \times {\mathscr{P}}_{Q},$ since
each $R_{q}(\mathbf{p})$ is a continuously differentiable function of $%
\mathbf{p}$ and the projector operator is continuous as well \cite[Prop. 3.2c%
]{Bertsekas Book-Parallel-Comp}.}\vspace{-0.3cm}
\begin{equation}
\mathbf{p}_{q}^{\star }=\mathbf{T}_{\mathbf{G}q}(\mathbf{p}^{\star })%
\triangleq%
\left[ \mathbf{p}_{q}^{\star }+\beta \mathbf{G}_{q}^{-1}%
\nabla%
_{q}R_{q}(\mathbf{p}_{q}^{\star },\mathbf{p}_{-q}^{\star })\right] _{{%
\mathscr{P}}_{q}}^{\mathbf{G}_{q}},\quad \forall q\in \Omega ,
\label{GP_mapping}
\end{equation}%
where $\left[ \mathbf{\cdot }\right] _{{\mathscr{P}}_{q}}^{\mathbf{G}_{q}}$
is the Euclidean projection on ${\mathscr{P}}_{q}$ with respect to the
vector norm $\left\Vert \mathbf{x}\right\Vert _{\mathbf{G}_{q}}%
\triangleq%
(\mathbf{x}^{T}\mathbf{G}_{q}\mathbf{x})^{1/2}.$
\end{lemma}

Given Lemma \ref{Lemma:GP_Mapping}, to reach the Nash equilibria of ${%
{\mathscr{G}}%
,}$ it is natural to employ an iterative algorithm, based either on Jacobi
or Gauss-Seidel schemes, using as best response for each user the mapping in
(\ref{GP_mapping}). Specifically, if the mapping in (\ref{GP_mapping}) is
used in the Jacobi scheme, we obtain the \emph{simultaneous} Iterative
Gradient Projection Algorithm (IGPA), as described in Algorithm \ref%
{SGPA_algo}.\bigskip

\begin{algo}{\hspace{0.6cm} Simultaneous Iterative Gradient Projection Algorithm}
SSet $\mathbf{p}_{q}^{(0)}=$ any feasible power allocation, $\forall q\in
\Omega,$ and $\beta>0$;\vspace{0.1cm}\newline
\texttt{for} $n=0:\mathrm{Nit}$
\begin{equation}
\,\,\,\,\mathbf{p}_{q}^{(n+1)}=\left[ \mathbf{p}_{q}^{(n)}+\beta \mathbf{G}_{q}^{-1}%
\nabla%
_{q}R_{q}(\mathbf{p}^{(n)})\right] _{{\mathscr{P}}%
_{q}}^{\mathbf{G}_{q}},\hspace{1cm}\forall
q\in\Omega, \label{SGPA}
\end{equation}%
\texttt{end}
\label{SGPA_algo}
\end{algo}
\bigskip
The sequential update of the strategies from the players can be easily
obtained from (\ref{GP_mapping}) by using the Gauss-Seidel scheme, and
provides the \emph{sequential }IGPA, as given in Algorithm \ref{GPA_algo}.\bigskip

\begin{algo}{Sequential Iterative Gradient Projection Algorithm}
SSet $\mathbf{p}_{q}^{(0)}=$ any
feasible power allocation, $\forall q\in
\Omega$, and $\beta>0$; \newline\texttt{for} $n=0:\mathrm{Nit}$ \newline
\begin{equation}
\,\mathbf{p}_{q}^{(n+1)}=\left\{
\begin{array}
[c]{ll}%
\left[
\mathbf{p}_{q}^{(n)}+\beta
\mathbf{G}_{q}^{-1} \nabla_{q}R_{q}\left( \mathbf{p}^{(n)}\right) \right] _{{\mathscr{P}}_{q}}^{\mathbf{G}_{q}}, & \text{if
}(n+1)\,\text{mod}\,Q=q,\\
\mathbf{p}_{q}^{(n)}, & \text{otherwise},
\end{array}
\right.  \hspace{0.2cm} \forall q\in\Omega;\label{Sequential-GPA}
\end{equation}
\newline
\texttt{end}\label{GPA_algo}
\end{algo}
\bigskip
The positive constant $\beta $ and the set of (positive definite) matrices $%
\{\mathbf{G}_{q}\}_{q\in \Omega }$ are free parameters that affect the
convergence property of the algorithms \cite%
{Bertsekas Book-Parallel-Comp}. Sufficient conditions for the convergence of
both sequential and simultaneous IGPAs are given in Appendix \ref{proof
Theorem Theo-SGPA}.\bigskip

\noindent \textbf{Remark 7 $-$ Computation of the projection.} According to the
best response mapping defined in (\ref{SGPA}) and (\ref{Sequential-GPA}),
both Algorithms \ref{SGPA_algo} and \ref{GPA_algo} require, at each
iteration, the computation of the Euclidean projection $\left[ \mathbf{\cdot
}\right] _{{\mathscr{P}}_{q}}^{\mathbf{G}_{q}}$ on the feasible strategy set
${\mathscr{P}}_{q}$ given in (\ref{admissible strategy set}), with respect
to the norm $\left\Vert \mathbf{\cdot }\right\Vert _{\mathbf{G}_{q}}$. $\ $%
For any given $\beta $, $\{\mathbf{G}_{q}\}_{q\in \Omega },$ and $\mathbf{p}%
=[\mathbf{p}_{1}^{T},\ldots ,\mathbf{p}_{Q}^{T}]^{T}\in {\mathscr{P},}$ the
projections in (\ref{SGPA}) and (\ref{Sequential-GPA}), written as
\begin{equation}
\mathbf{T}_{\mathbf{G}q}(\mathbf{p})=\left[ \mathbf{p}_{q}-\beta \mathbf{G}%
_{q}^{-1}f_{q}(\mathbf{p})\right] _{{\mathscr{P}}_{q}}^{\mathbf{G}%
_{q}},\quad \text{with}\quad \text{ }f_{q}(\mathbf{p})=-%
\nabla%
_{q}R_{q}(\mathbf{p}),  \label{T_g_p_0}
\end{equation}%
can be computed solving the following convex quadratic programming:%
\begin{equation}
\begin{array}{ll}
\limfunc{minimize}\limits_{\widetilde{\mathbf{p}}_{q}} & \dfrac{1}{2}%
\widetilde{\mathbf{p}}_{q}^{T}\mathbf{M}_{q}\widetilde{\mathbf{p}}_q+%
\widetilde{\mathbf{p}}_{q}^{T}\mathbf{q}_{q} \\
\limfunc{subject}\limfunc{to} & \widetilde{\mathbf{p}}_{q}\in {\mathscr{P}}%
_{q},%
\end{array}
\label{convex_QP}
\end{equation}%
where
\begin{equation}
\mathbf{M}_{q}%
\triangleq%
(1/\beta )\mathbf{G}_{q},\quad \text{and}\quad \text{ }\mathbf{q}_{q}=%
\mathbf{q}_{q}(\mathbf{p})%
\triangleq%
f_{q}(\mathbf{p})-(1/\beta )\mathbf{G}_{q}\mathbf{p}_{q},
\label{def:M_and_q}
\end{equation}%
with $f_{q}(\mathbf{p})$ given in (\ref{T_g_p_0}). Observe that in the
special case of $\mathbf{G}_{q}=\mathbf{I},$ the mapping $\mathbf{T}_{%
\mathbf{G}q}(\mathbf{p})$ in (\ref{T_g_p_0}) becomes the classical Euclidean
projection on the set ${\mathscr{P}}_{q},$ that can be efficiently computed
as a waterfilling solution, as shown in Section \ref{Sec:single-user WF}
(cf. Lemma \ref{Euclidean Projection}).

Interestingly, to compute the projection in (\ref{T_g_p_0}), a variety of
alternative algorithms can be obtained, interpreting $\mathbf{T}_{\mathbf{G}%
q}(\mathbf{p})$ in (\ref{T_g_p_0}) as the unique solution to a proper Linear
Variational Inequality (LVI) problem \cite{Bertsekas Book-Parallel-Comp}, as
we show next. \ Using the scaled projection Theorem \cite[Prop. 3.7(b)]%
{Bertsekas Book-Parallel-Comp},\footnote{%
The scaled projection theorem says that, given some $\mathbf{x}\in
\mathbb{R}
^{n}$ and a convex set $\mathcal{X}\subseteq
\mathbb{R}
^{n},$ a vector $\mathbf{z}\in \mathcal{X}$ is equal to $\left[ \mathbf{x}%
\right] _{\mathcal{X}}^{\mathbf{G}}$ if and only if $(\mathbf{y}-\mathbf{z}%
)^{T}\mathbf{G}(\mathbf{y}-\mathbf{z})\leq 0,$ $\forall \mathbf{y}\in
\mathcal{X},$ where $\left[ \mathbf{x}\right] _{\mathcal{X}}^{\mathbf{G}}$
denotes the Euclidean projection of $\mathbf{x}$ on $\mathcal{X}$ with
respect to the norm $\left\Vert \mathbf{\cdot }\right\Vert _{\mathbf{G}}.$}
one can find that the projection $\mathbf{T}_{\mathbf{G}q}(\mathbf{p})$ in (%
\ref{T_g_p_0}) can be equivalently defined as the unique vector $\mathbf{p}%
_{q}^{\star }\in {\mathscr{P}}_{q}$ such that%
\begin{equation*}
(\mathbf{y}_{q}-\mathbf{p}_{q}^{\star })^{T}\mathbf{G}_{q}\left( \mathbf{p}%
_{q}-\beta \mathbf{G}_{q}^{-1}f_{q}(\mathbf{p})-\mathbf{p}_{q}^{\star
}\right) \leq 0,\quad \forall \mathbf{y}_{q}\in {\mathscr{P}}_{q},
\end{equation*}%
which, since $\beta >0,$ can be rewritten as%
\begin{equation}
\text{(LVI)\qquad \qquad }(\mathbf{y}_{q}-\mathbf{p}_{q}^{\star })^{T}\left(
\mathbf{M}_{q}\mathbf{p}_{q}^{\star }+\mathbf{q}_{q}\right) \geq 0,\quad
\forall \mathbf{y}_{q}\in {\mathscr{P}}_{q},  \label{LVI}
\end{equation}%
where $\mathbf{M}_{q}$ and $\mathbf{q}_{q}$ are defined in (\ref{def:M_and_q}%
). Inequality in (\ref{LVI}) still defines a variational inequality problem
(see (\ref{VI})), but computationally simpler than the original one given in
(\ref{KKT-NE}), as the function $\mathbf{M}_{q}\mathbf{p}_{q}^{\star }+%
\mathbf{q}_{q}(\mathbf{p})$ in (\ref{LVI}), for any given $\mathbf{p},$ is
linear in $\mathbf{p}_{q}^{\star }.$ Observe that, since $\mathbf{G}_{q}$
(and thus $\mathbf{M}_{q}$) is positive definite, the (unique) solution $%
\mathbf{p}_{q}^{\star }$ in (\ref{LVI}) is well-defined, as LVI in (\ref{LVI}%
) admits a unique solution \cite[Prop. 5.5]{Bertsekas Book-Parallel-Comp}. A
variety of algorithms, known in the literature as linearized algorithms can
be used to efficiently solve the LVI in (\ref{LVI}). The interested reader
may refer to \cite{Facchinei, Nagurney, Pang} for a broad overview of these
algorithms.\bigskip

\noindent \textbf{Remark 8 $-$ Distributed nature of the algorithms. }%
Interestingly, both IGPAs keep the most appealing features of IWFAs, namely
its low-complexity distributed nature. In fact, as in IWFAs, also in IGPAs
each user needs only to locally measure the PSD of the overall interference received
from the other users and project a vector that depends on this interference
(i.e., $\nabla _{q}R_{q}(\mathbf{p})$) onto its own feasible set.

\bigskip

\noindent \textbf{Numerical Example. }As an example, in Figure \ref%
{fig:SIWFA_vs_IWFA}, we compare the performance of the simultaneous IGPA
with the simultaneous IWFA, in terms of convergence speed. We consider a
network composed of $Q=35$ active users and compare the rate evolution of $3$
out of $35$ links as a function of the iteration index $n,$ as defined in
Algorithms \ref{SIWFA_algo} and \ \ref{SGPA_algo}. Interestingly, the
simultaneous IGPA shows similar convergence speed than simultaneous IWFA.
Thus, it can be used as a valid alternative to the simultaneous IWFA.

\begin{figure}[tbp]
\vspace{-0.8cm}\centering
{\includegraphics[height=8cm]{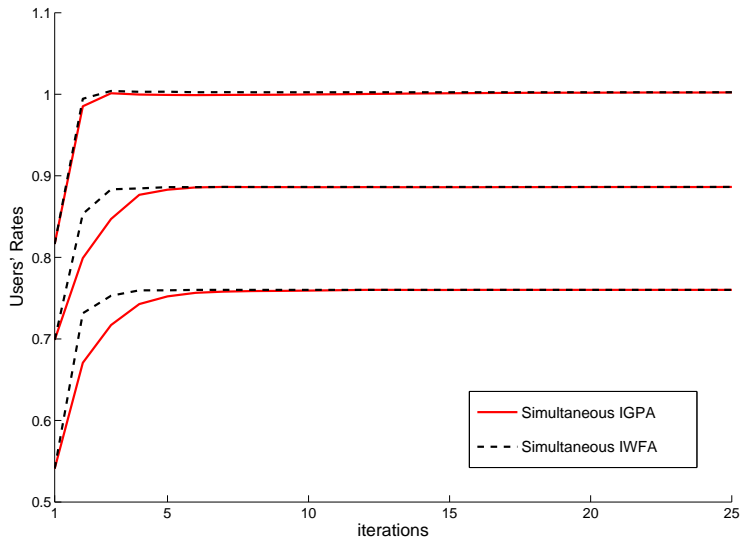} \label{fig:SGPA_vs_IWFA}}%
\vspace{-0.6cm} \vspace{-0.4cm}
\caption{{\protect\scriptsize Users' rates versus iterations; simultaneous
IGPA (solid line curves) and simultaneous IWFA (dashed-line curves), $Q=35$,
$\protect\gamma =2.5,$ $d_{rq}$ $=d_{qr},$ $d_{rr}=d_{qq}=1$, $P_q=P_r$, $%
\Gamma_{q} =1$, $P_q/\protect\sigma_q^2=7$dB, $P_r/(\protect\sigma_q^2
d_{rq}^\protect\gamma)=3$dB, $\forall r,q\in \Omega$.}}
\label{fig:SIWFA_vs_IWFA}
\end{figure}
\vspace{-0.3cm}

\section{Conclusions\label{Conclusions}}

In this two-part paper, we have formulated the problem of finding the
optimal  linear transceivers in a multipoint-to-multipoint
wideband network, as a strategic non-cooperative game. We first considered
the theoretical problem of maximizing mutual information on each link, given
constraints on the spectral mask and transmit power. Then, to accommodate
for practical implementation aspects, we focused on the competitive maximization
of the transmission rate on each link, using finite order constellations,
under the same constraints as above plus a constraint on the average error
probability. In Part I of the paper we fully characterized both games by
providing a unified expression for the optimal structure of the linear
transceivers and deriving conditions for the uniqueness of the NE. In this
Part II, we have focused on how to reach these equilibria using totally
decentralized algorithms. We have proposed and analyzed alternative
distributed iterative algorithms along with their convergence conditions,
namely: 1) the sequential IWFA, which is a generalization of the well-known
(sequential) iterative waterfilling algorithm proposed by Yu et. al. to the
case where spectral mask constraints are incorporated in the optimization;
and 2) the simultaneous IWFA, which has been shown to converge faster than
the sequential IWFA; 3) The sequential and simultaneous IGPAs, which are
based on the gradient projection best response, and establish an interesting
link between the Nash equilibria of the game and the solutions to the
corresponding variational inequality problem. Interestingly, the
simultaneous IGPA has been shown to have approximately the same convergence
speed and computational complexity of the simultaneous IWFA, and thus it can
be a valid alternative to the algorithms based on the waterfilling
solutions. We have derived the sufficient conditions
for the \emph{global }convergence of all the proposed algorithms that,
although proved under stronger constraints (e.g., the additional spectral
mask constraint), have broader validity than the convergence conditions
known in the current literature for the sequential IWFA proposed by Yu et al.

We are currently investigating the extension of the proposed algorithms to
the case in which the updating strategies are
performed in a totally asynchronous way \cite{SPAWC-AIWFA}. The other major extension that
needs to be addressed is the situation where the channels and interference
covariance matrices are known only within an inevitable estimation error.

\vspace{-0.3cm}

\section{Appendix}

\appendix

\section{Proof of Lemma \protect\ref{Euclidean Projection}}

\label{proof_of_Euclidean Projection}

First of all, observe that the objective function of the convex problem (\ref%
{convex-projection}) is coercive on the feasible set \cite%
{Bazaraa-Sherali-Shetty}. Hence, a solution for the problem (\ref%
{convex-projection}) exists \cite{Bazaraa-Sherali-Shetty}. Since problem (%
\ref{convex-projection}) satisfies Slater's condition \cite{Boyd,
Bazaraa-Sherali-Shetty}, the Karush-Kuhn-Tucker (KKT) conditions are both
necessary and sufficient for the optimality. The Lagrangian corresponding to
the constrained convex problem (\ref{convex-projection}) is
\begin{equation}
\mathcal{L}=\dfrac{1}{2}\dsum_{k=1}^{N}\mathbf{(}x_{k}\mathbf{\mathbf{+}}%
x_{0,k}\mathbf{)}^{2}-\dsum_{k=1}^{N}\nu _{k}x_{k}-\widetilde{\mu }\left(
\frac{1}{N}\sum_{k=1}^{N}x_{k}-1\right) +\dsum_{k=1}^{N}\lambda _{k}\left(
x_{k}-p^{\max }(k)\right) ,
\end{equation}%
and the KKT conditions are:
\begin{equation}
\begin{array}{l}
x_{k}\mathbf{\mathbf{+}}x_{0,k}-\nu _{k}+\lambda _{k}-\mu =0, \\
\nu _{k}\geq 0,\quad x_{k}\geq 0,\quad \nu _{k}x_{k}=0, \\
\lambda _{k}\geq 0,\quad x_{k}\leq p^{\max }(k),\quad \lambda _{k}\left(
x_{k}-p^{\max }(k)\right) =0, \\
\dfrac{1}{N}\sum\limits_{k=1}^{N}x_{k}=1,%
\end{array}%
\quad k\in \{1,\ldots ,N\},  \label{KKT}
\end{equation}%
with $x_{0,k}\geq 0$, $\forall k\in \{1,\ldots ,N\}$ and $\mu
\triangleq%
\widetilde{\mu }/N.$

Observe that, if $(1/N)\sum_{k=1}^{N}p^{\max }(k)<1$ or $p^{\max }(k)<0$ for
some $k,$ then the problem is infeasible; if $(1/N)\sum_{k=1}^{N}p^{\max
}(k)=1,$ then the problem admits the trivial solution $x_{k}=$ $p^{\max
}(k), $ $\forall k;$ if $p^{\max }(k)=0$ for some $k,$ then $x_{k}=0.$ Here after, we thus assume that all the subcarrier indices corresponding to the
zero-valued $p^{\max }(k)^{\text{'}}$s have been removed and $%
(1/N)\sum_{k=1}^{N}p^{\max }(k)>1$ (to avoid the trivial solution).

First of all, observe that $\mu >0$ (in fact, $\mu \leq 0$ is not
admissible, since the constraint $(1/N)\sum_{k=1}^{N}x_{k}=1$ necessarily
implies $x_{k}\mathbf{\mathbf{=-}}x_{0,k}-\lambda _{k}+\mu >0,$ for at least
one $k$). If $x_{k}=0,$ since $\nu _{k}\geq 0$ and (by the complementary
slackness condition) $\lambda _{k}=0,$ then $\mu -x_{0,k}=-\nu _{k}\leq 0.$
If $\ 0<x_{k}<p^{\max }(k),$ then $\nu _{k}=0$ and $\lambda _{k}=0;$ which
provides $x_{k}=\mu -x_{0,k},$ (observe that $0<\mu -x_{0,k}<p^{\max }(k)$).
Finally, if $x_{k}=p^{\max }(k),$ then $\nu _{k}=0$ and $\lambda _{k}\geq 0,$
which implies $\mu -x_{0,k}\geq p^{\max }(k).$ Since, for each $k,$ the
values of the admissible solution induce a partition on the set of the $\mu $
values, the solution can be written as
\begin{equation*}
x_{k}=\left[ \mu -x_{0,k}\right] _{0}^{p^{\max }(k)}=\left\{
\begin{array}{ll}
0, & \text{if }\mu -x_{0,k}\leq 0, \\
\mu -x_{0,k}, & \text{if }0<\mu -x_{0,k}<p^{\max }(k), \\
p^{\max }(k), & \text{if }\mu -x_{0,k}\geq p^{\max }(k),%
\end{array}%
\right. \quad k\in \{1,\ldots ,N\},
\end{equation*}%
where $\mu $ is chosen so that $(1/N)\sum_{k=1}^{N}\left[ \mu -x_{0,k}\right]
_{0}^{p^{\max }(k)}=1$.\hspace{\fill}\rule{1.5ex}{1.5ex} \vspace{-0.3cm}

\section{Properties of Waterfilling Projection}

\label{Appendix WF-Projection}

\label{Appendix_WF_project} First of all, it is convenient to rewrite the
waterfilling operator in (\ref{WF-projection}) as
\begin{equation}
\mathsf{WF}%
_{q}\left( \mathbf{p}_{-q}\right) =\left[ -\mathbf{\sigma }%
_{q}-\sum_{\,r\neq q}\mathbf{H}_{rq}\mathbf{p}_{r}\right] _{{\mathscr{P}}%
_{q}},  \label{WF-op2}
\end{equation}%
where%
\begin{equation}
\begin{array}{l}
\mathbf{H}_{rq}%
\triangleq%
\Gamma _{q}\limfunc{diag}\left( \dfrac{\left\vert H_{rq}(1)\right\vert ^{2}}{%
\left\vert H_{qq}(1)\right\vert ^{2}},\ldots ,\dfrac{\left\vert
H_{rq}(N)\right\vert ^{2}}{\left\vert H_{qq}(N)\right\vert ^{2}}\right)
,\quad \bigskip \vspace{-0.15cm} \\
\ \ \mathbf{\sigma }_{q}%
\triangleq%
\Gamma _{q}\left[ 1/\left\vert H_{qq}(1)\right\vert ^{2},\ldots
,1/\left\vert H_{qq}(N)\right\vert ^{2}\right] ^{T},%
\end{array}%
\quad \forall r\neq q,q\in \Omega ,  \label{def:H_rq}
\end{equation}%
and ${\mathscr{P}}_{q}$ is defined in (\ref{admissible strategy set}).

Building on (\ref{WF-op2}), we derive now a key property of the waterfilling
operator that will be fundamental in proving Theorems \ref{Theo-IWFA} and %
\ref{Theo-SIWFA}. To this end, we introduce the following mapping. For
technical reasons, we first define
\begin{equation}
\widetilde{p}_{q}^{\max }(k)%
\triangleq%
\left\{
\begin{array}{ll}
p_{q}^{\max }(k), & \text{if }k\in \mathcal{D}_{q}, \\
0, & \text{otherwise,}%
\end{array}%
\right.
\end{equation}%
where $\mathcal{D}_{q}$ is given in (\ref{D_q}), and introduce the
admissible set ${\mathscr{P}}^{\text{eff}}={\mathscr{P}}_{1}^{\text{eff}%
}\times \cdots \times {\mathscr{P}}_{Q}^{\text{eff}}\subseteq {\mathscr{P},}$
where ${\mathscr{P}}_{q}^{\text{eff}}$ is the subset of ${\mathscr{P}}_{q}$
containing all the feasible power allocations of user $q$, with zero power
over the carriers that user $q$ would never use, for the given power budget
and interference level, in any of its waterfilling solutions (\ref{WF_mask}%
), against any admissible strategy of the others:%
\begin{equation}
\begin{array}{l}
{\mathscr{P}}_{q}^{\text{eff}}%
\triangleq%
\left\{ \mathbf{p}_{q}\in {\mathscr{P}}_{q}\text{ with }p_{q}(k)=0\text{ }%
\forall k\notin \mathcal{D}_{q}\right\} \\
\qquad \ =\left\{ \mathbf{p}_{q}\in \mathcal{\ \mathbb{R}}^{N}:\dfrac{1}{N}\
\dsum\limits_{k=1}^{N}p_{q}(k)=1,\text{ }0\leq p_{q}(k)\leq \widetilde{p}%
_{q}^{\max }(k),\text{ \ }\forall k\in \{1,\ldots ,N\}\right\} ,%
\end{array}
\label{X_q set}
\end{equation}%
where the second equality in (\ref{X_q set}) follows from the properties of
the waterfilling solution (\ref{WF_mask}) (cf. Appendix \ref%
{proof_of_Euclidean Projection}). Observe that, because of (\ref{X_q set}),
the game does not change if we use ${\mathscr{P}}^{\text{eff}}$ instead of
the original ${\mathscr{P}}$. For any given $\{\alpha _{q}\}_{q\in \Omega }$
with $\alpha _{q}\in \lbrack 0,1),$ let $\mathbf{T}(\mathbf{p})=(\mathbf{T}%
_{q}(\mathbf{p}))_{q\in \Omega }:{\mathscr{P}}^{\text{eff}}{\mapsto %
\mathscr{P}}^{\text{eff}}$ be the mapping defined, for each $q,$ as%
\begin{align}
\mathbf{T}_{q}(\mathbf{p})&
\triangleq%
\alpha _{q}\ \mathbf{p}_{q}+(1-\alpha _{q})\left[ -%
\boldsymbol{\sigma}_q%
-\sum_{\,r\neq q}\mathbf{H}_{rq}\mathbf{p}_{r}\right] _{{\mathscr{P}}_{q}^{%
\text{eff}}}  \notag \\
& =\alpha _{q}\ \mathbf{p}_{q}+(1-\alpha _{q})\left[ -%
\boldsymbol{\sigma}_q%
-\sum_{\,r\neq q}\mathbf{H}_{rq}\mathbf{p}_{r}\right] _{{\mathscr{P}}%
_{q}},\quad \mathbf{p}\in {\mathscr{P}}^{\text{eff}}\text{,\quad }\alpha
_{q}\in \lbrack 0,1),  \label{Mapping_T}
\end{align}%
where the second equality follows from (\ref{X_q set})$.$ Observe that the
operator in (\ref{Mapping_T}) is indeed a mapping from ${\mathscr{P}}^{\text{%
eff}}$ to ${\mathscr{P}}^{\text{eff}}$, due to the convexity of ${\mathscr{P}%
}^{\text{eff}}$. Moreover, it follows from (\ref{WF-op2}) that all the Nash
equilibria $\mathbf{p}^{\star }%
\triangleq%
\left( \mathbf{p}_{q}^{\star }\right) _{q\in \Omega }$ of game ${%
{\mathscr{G}}%
}$ in (\ref{Rate Game}) (see (\ref{fixed-point_NE})) satisfy the following
set of equations%
\begin{equation}
\mathbf{p}_{q}^{\star }=\left[ -%
\boldsymbol{\sigma}_q%
-\sum_{\,r\neq q}\mathbf{H}_{rq}\mathbf{p}_{r}^{\star }\right] _{{\mathscr{P}%
}_{q}^{\text{eff}}},\quad \forall q\in \Omega ;
\end{equation}%
which correspond to the fixed points in ${\mathscr{P}}^{\text{eff}}$ of the
mapping\ $\mathbf{T}$ defined in (\ref{Mapping_T}). Hence, the existence of
at least one fixed point for $\mathbf{T}$ is guaranteed by the existence of
a NE for game ${%
{\mathscr{G}}%
}$ {\cite[Theorem 2]{Scutari-Part I}}.

Before proving the main property of the mapping $\mathbf{T,}$ we need the
following intermediated definitions. Given $\mathbf{T}$ in (\ref{Mapping_T})
and some $\mathbf{w}%
\triangleq%
[w_{q},\ldots ,w_{Q}]^{T}>\mathbf{0}$, let $\left\Vert \cdot \right\Vert _{2,%
\text{block}}^{\mathbf{w}}$ denote the (vector) block-maximum norm$,$
defined as \cite{Bertsekas Book-Parallel-Comp}
\begin{equation}
\left\Vert \mathbf{T}(\mathbf{p})\right\Vert _{2,\text{block}}^{\mathbf{w}}%
\triangleq%
\max_{q\in \Omega }\frac{\left\Vert \mathbf{T}_{_{q}}(\mathbf{p})\right\Vert
_{2}}{w_{q}},\text{ }\quad  \label{block_max_weight_norm}
\end{equation}%
where $\left\Vert \mathbf{\cdot }\right\Vert _{2}$ is the Euclidean norm.
Let $\left\Vert \mathbf{\cdot }\right\Vert _{\infty ,\text{vec}}^{\mathbf{w}%
} $ be the \emph{vector }weighted maximum norm$,$ defined as \cite{Horn85}
\begin{equation}
\left\Vert \mathbf{x}\right\Vert _{\infty ,\text{vec}}^{\mathbf{w}}%
\triangleq%
\max_{q\in \Omega }\frac{\left\vert x_{q}\right\vert }{w_{q}},\quad \mathbf{%
w>0,}\text{ }\quad \mathbf{x\in
\mathbb{R}
}^{Q},  \label{weighted_infinity_vector_norm}
\end{equation}%
and let $\left\Vert \mathbf{\cdot }\right\Vert _{\infty ,\text{mat}}^{%
\mathbf{w}}$ denote the \emph{matrix} norm induced by $\left\Vert \cdot
\right\Vert _{\infty ,\text{vec}}^{\mathbf{w}},$ defined as \cite{Horn85}
\begin{equation}
\left\Vert \mathbf{A}\right\Vert _{\infty ,\text{mat}}^{\mathbf{w}}%
\triangleq%
\max_{q}\frac{1}{w_{q}}\dsum\limits_{r=1}^{Q}[\mathbf{A}]_{qr}\text{ }w_{r},%
\text{ }\quad \mathbf{A\in
\mathbb{R}
}^{Q\times Q}.  \label{H_max_weight_norm}
\end{equation}%
Finally, define $\left\Vert \mathbf{\cdot }\right\Vert _{2,\mathcal{D}_{q}}$
as
\begin{equation}
\left\Vert \mathbf{x}\right\Vert _{2,\mathcal{D}_{q}}%
\triangleq%
\left( \sum_{k\in \mathcal{D}_{q}}\left( x_{k}\right) ^{2}\right)
^{1/2},\quad \mathbf{x\in
\mathbb{R}
}^{N},  \label{L2_Dq_norm}
\end{equation}%
with $\mathcal{D}_{q}$ defined in (\ref{D_q}). Observe that $\left\Vert
\mathbf{\cdot }\right\Vert _{2,\mathcal{D}_{q}}$ is not a vector norm (as
does not satisfy the positivity property), \ but it is a vector seminorm \cite%
{Horn85}.

The mapping $\mathbf{T}$ in (\ref{Mapping_T}) is said to be a \textit{%
block-contraction} with modulus $\beta ,$ with respect to the norm $%
\left\Vert \cdot \right\Vert _{2,\text{block}}^{\mathbf{w}}$\ in (\ref%
{block_max_weight_norm}), if there exists $\beta \in \lbrack 0,1)$ such that
\cite[Sec. 3.1.2]{Bertsekas Book-Parallel-Comp}%
\begin{equation}
\left\Vert \mathbf{T}(\mathbf{p}^{(1)})-\mathbf{T}(\mathbf{p}%
^{(2)})\right\Vert _{2,\text{block}}^{\mathbf{w}}\leq \beta \left\Vert
\mathbf{p}^{(1)}-\mathbf{p}^{(2)}\right\Vert _{2,\text{block}}^{\mathbf{w}%
},\quad \forall \mathbf{p}^{(1)},\mathbf{p}^{(2)}\in {\mathscr{P}}_{q}^{%
\text{eff}}{.}  \label{pseudo-contraction_def}
\end{equation}%
We provide now some interesting properties for the mapping $\mathbf{T}$ in (%
\ref{Mapping_T}), that will be instrumental to study the convergence of both
sequential and simultaneous IWFAs.

\begin{lemma}[Nonexpansive property of the waterfilling mapping]
\label{NonExpansive-Lemma}Given $\mathcal{D}_{q},$ ${\mathscr{P}}_{q}^{\text{%
eff}},$ and $\left\Vert \mathbf{\cdot }\right\Vert _{2,\mathcal{D}_{q}}$
defined in (\ref{D_q}), (\ref{X_q set}) and (\ref{L2_Dq_norm}),
respectively, let $\left[ \cdot \right] _{{\mathscr{P}}_{q}^{\text{eff}%
}}^{\left\Vert \cdot \right\Vert }$ denote the projector operator onto the
convex set ${\mathscr{P}}_{q}^{\text{eff}}$ with respect to the vector norm $%
\left\Vert \cdot \right\Vert $. Then, \ $\left[ \cdot \right] _{{\mathscr{P}}%
_{q}^{\text{eff}}}^{\left\Vert \cdot \right\Vert _{2}}$\ satisfies the
following nonexpansive property:%
\begin{equation}
\left\Vert \left[\,\mathbf{x}\,\right] _{{\mathscr{P}}_{q}^{\text{eff}%
}}^{\left\Vert \cdot \right\Vert _{2}}-\left[\,\mathbf{y}\,\right] _{{%
\mathscr{P}}_{q}^{\text{eff}}}^{\left\Vert \cdot \right\Vert
_{2}}\right\Vert _{2,\mathcal{D}_{q}}\leq \left\Vert \mathbf{x}-\mathbf{y}%
\right\Vert _{2,\mathcal{D}_{q}},\qquad \forall \mathbf{x,y\in
\mathbb{R}
}_{+}^{N}.  \label{NonExpansive_ineq}
\end{equation}
\end{lemma}

\begin{proof}
For any given $\varepsilon >0,$ let $\left\Vert \cdot \right\Vert _{2,%
\mathcal{D}_{q}}^{\varepsilon }$ denote the weighted vector norm (derived
from an inner product \cite{Horn85}), defined as
\begin{equation}
\left\Vert \mathbf{x}\right\Vert _{2,\mathcal{D}_{q}}^{\varepsilon }%
\triangleq%
\left( \sum_{k\in \mathcal{D}_{q}}\left( x_{k}\right) ^{2}+\varepsilon
\sum_{k\notin \mathcal{D}_{q}}\left( x_{k}\right) ^{2}\right) ^{1/2},\quad
\mathbf{x\in
\mathbb{R}
}^{N},\quad \varepsilon >0.  \label{L2_Dq_epsilin}
\end{equation}%
Then the projector $\left[\, \mathbf{\cdot }\,\right] _{{\mathscr{P}}_{q}^{\text{%
eff}}}^{\left\Vert \mathbf{\cdot }\right\Vert _{2,\mathcal{D}%
_{q}}^{\varepsilon }}$ \ satisfies the following inequality:%
\begin{equation}
\left\Vert \left[\,\mathbf{x}\,\right] _{{\mathscr{P}}_{q}^{\text{eff}%
}}^{\left\Vert \mathbf{\cdot }\right\Vert _{2}}-\left[\,\mathbf{y}\,\right] _{{%
\mathscr{P}}_{q}^{\text{eff}}}^{\left\Vert \mathbf{\cdot }\right\Vert
_{2}}\right\Vert _{2,\mathcal{D}_{q}}^{\varepsilon }=\left\Vert \left[\,
\mathbf{x}\,\right] _{{\mathscr{P}}_{q}^{\text{eff}}}^{\left\Vert \mathbf{%
\cdot }\right\Vert _{2,\mathcal{D}_{q}}^{\varepsilon }}-\left[\,\mathbf{y}\,
\right] _{{\mathscr{P}}_{q}^{\text{eff}}}^{\left\Vert \mathbf{\cdot }%
\right\Vert _{2,\mathcal{D}_{q}}^{\varepsilon }}\right\Vert _{2,\mathcal{D}%
_{q}}^{\varepsilon }\leq \left\Vert \mathbf{x}-\mathbf{y}\right\Vert _{2,%
\mathcal{D}_{q}}^{\varepsilon },\quad \forall \mathbf{x,y\in
\mathbb{R}
}_{+}^{N},\text{ }\forall \varepsilon >0,  \label{L2_D_2_ineq}
\end{equation}%
where the equality in (\ref{L2_D_2_ineq}) follows from $\left[\,\mathbf{x}
_{0}\,\right] _{{\mathscr{P}}_{q}^{\text{eff}}}^{\left\Vert \mathbf{\cdot }%
\right\Vert _{2}}=\left[\,\mathbf{x}_{0}\,\right] _{{\mathscr{P}}_{q}^{\text{%
eff}}}^{\left\Vert \mathbf{\cdot }\right\Vert _{2,\mathcal{D}%
_{q}}^{\varepsilon }},$ due to the equivalence between the optimization
problem (\ref{convex-projection}) and the same problem where the original
objective function is replaced by $(\left\Vert \mathbf{x}-(-\mathbf{x}%
_{0})\right\Vert _{2,\mathcal{D}_{q}}^{\varepsilon})^2=\sum_{k\in \mathcal{%
D}_{q}}\left( x_{k}+x_{0,k}\right) ^{2}+\varepsilon \sum_{k\notin \mathcal{D}%
_{q}}\left( -x_{0,k}\right) ^{2}$ (since any $\mathbf{x\in }{\mathscr{P}}%
_{q}^{\text{eff}}$ is such that  $x_{k}=0,\ \forall k\notin \mathcal{D}_{q}$%
); and the inequality in (\ref{L2_D_2_ineq}) represents the nonexpansion
property of the projector $\left[\, \cdot \,\right] _{{\mathscr{P}}_{q}^{\text{%
eff}}}^{\left\Vert \mathbf{\cdot }\right\Vert _{2,\mathcal{D}%
_{q}}^{\varepsilon }}$ in the norm $\left\Vert \mathbf{\cdot }\right\Vert
_{2,\mathcal{D}_{q}}^{\varepsilon }$ \cite[Prop. 3.2(c)]{Bertsekas
Book-Parallel-Comp}.\footnote{%
Observe that the nonexpansive property of the projector, usually given in
the Euclidean norm, is preserved in any vector norm (derived from an inner
product) used to define the projection.}

Since $\left\Vert \cdot  \right\Vert _{2,%
\mathcal{D}_{q}}^{\varepsilon }$ is a continuous function of $\varepsilon >0,
$ taking in (\ref{L2_D_2_ineq}) the limit as $\varepsilon \rightarrow 0,$
and using%
\begin{equation*}
\lim_{\varepsilon \rightarrow 0}\left\Vert \cdot \right\Vert _{2,\mathcal{D}%
_{q}}^{\varepsilon }=\left\Vert \cdot \right\Vert _{2,\mathcal{D}_{q}},
\end{equation*}%
we obtain the desired inequality, as stated in (\ref{NonExpansive_ineq}).
\end{proof}

\begin{proposition}[Contraction property of mapping $\mathbf{T}$]
\label{Proposition:block-contraction}Given $\mathbf{w}%
\triangleq%
[w_{1},\ldots ,w_{Q}]^{T}\mathbf{>0,}$ assume that the following condition
is satisfied:
\begin{equation}
\Vert \mathbf{H}^{\max }\Vert _{\infty ,\text{\emph{mat}}}^{\mathbf{w}}<1%
\mathbf{,}  \label{b_modulus_contraction}
\end{equation}%
where $\mathbf{H}^{\max }$ and $\left\Vert \cdot \right\Vert _{\infty ,\text{%
\emph{mat}}}^{\mathbf{w}}$ are defined\ in (\ref{Hmax}) and (\ref%
{H_max_weight_norm}), respectively. Then, the mapping $\mathbf{T}$ defined
in (\ref{Mapping_T})$\mathbf{\ }$is a \textit{block-contraction with modulus
}$\beta =\Vert \mathbf{H}^{\max }\Vert _{\infty ,\text{\emph{mat}}}^{\mathbf{%
w}},$ with respect to the block-maximum norm $\left\Vert \cdot \right\Vert
_{2,\text{\emph{block}}}^{\mathbf{w}}$ defined in (\ref{block_max_weight_norm}).
\end{proposition}

\begin{proof}
The proof consists in showing that, under (\ref{b_modulus_contraction}), the
mapping $\mathbf{T}$ satisfies (\ref{pseudo-contraction_def}), with $\beta
=\Vert \mathbf{H}^{\max }\Vert _{\infty }^{\mathbf{w}}.$ Given $\mathbf{p}%
^{(1)}=\left( \mathbf{p}_{q}^{(1)},\ldots ,\mathbf{p}_{Q}^{(1)}\right) \in {%
\mathscr{P}}^{\text{eff}}$ and $\mathbf{p}^{(2)}=\left( \mathbf{p}%
_{1}^{(2)},\ldots ,\mathbf{p}_{Q}^{(2)}\right) \in {\mathscr{P}}^{\text{eff}%
},$ define
\begin{equation}
e_{\mathbf{T}q}%
\triangleq%
\left\Vert \mathbf{T}_{q}\mathbf{(p}^{(1)}\mathbf{)}-\mathbf{T}_{q}\mathbf{(p%
}^{(2)}\mathbf{)}\right\Vert _{2}\mathbf{\quad }\text{and}\mathbf{\quad }%
e_{q}%
\triangleq%
\left\Vert \mathbf{p}_{q}^{(1)}-\mathbf{p}_{q}^{(2)}\right\Vert _{2},\quad
q\in \Omega .  \label{e_q_and_e_T_q}
\end{equation}%
Then, we have:%
\begin{align}
e_{\mathbf{T}q}& =\left\Vert \mathbf{T}_{q}\mathbf{(p}^{(1)}\mathbf{)}-%
\mathbf{T}_{q}\mathbf{(p}^{(2)}\mathbf{)}\right\Vert _{2} \\
& \leq \alpha _{q}\left\Vert \mathbf{p}_{q}^{(1)}-\mathbf{p}%
_{q}^{(2)}\right\Vert _{2}+(1-\alpha _{q})\left\Vert \left[ -%
\boldsymbol{\sigma}_q%
-\sum_{\,r\neq q}\mathbf{H}_{rq}\mathbf{p}_{r}^{(1)}\right] _{{\mathscr{P}}%
_{q}^{\text{eff}}}-\left[ -%
\boldsymbol{\sigma}_q%
-\sum_{\,r\neq q}\mathbf{H}_{rq}\mathbf{p}_{r}^{(2)}\right] _{{\mathscr{P}}%
_{q}^{\text{eff}}}\right\Vert _{2}  \label{Ineq_2} \\
& =\alpha _{q}e_{q}+(1-\alpha _{q})\left\Vert \left[ -%
\boldsymbol{\sigma}_q%
-\sum_{\,r\neq q}\mathbf{H}_{rq}\mathbf{p}_{r}^{(1)}\right] _{{\mathscr{P}}%
_{q}^{\text{eff}}}-\left[ -%
\boldsymbol{\sigma}_q%
-\sum_{\,r\neq q}\mathbf{H}_{rq}\mathbf{p}_{r}^{(2)}\right] _{{\mathscr{P}}%
_{q}^{\text{eff}}}\right\Vert _{2,\mathcal{D}_{q}}  \label{Ineq_3} \\
& \leq \alpha _{q}e_{q}+(1-\alpha _{q})\left\Vert \sum_{\,r\neq q}\mathbf{H}%
_{rq}\mathbf{p}_{r}^{(1)}-\sum_{\,r\neq q}\mathbf{H}_{rq}\mathbf{p}%
_{r}^{(2)}\right\Vert _{2,\mathcal{D}_{q}}  \label{ineq_6} \\
& =\alpha _{q}e_{q}+(1-\alpha _{q})\left\Vert \sum_{\,r\neq q}\overline{%
\mathbf{H}}_{rq}\left( \mathbf{p}_{r}^{(1)}-\mathbf{p}_{r}^{(2)}\right)
\right\Vert _{2}  \label{ineq_7} \\
& \leq \alpha _{q}e_{q}+(1-\alpha _{q})\sum_{\,r\neq q}\left( \max_{k}\left[
\overline{\mathbf{H}}_{rq}\right] _{kk}\right) \left\Vert \mathbf{p}%
_{r}^{(1)}-\mathbf{p}_{r}^{(2)}\right\Vert _{2}  \label{ineq_8} \\
& =\alpha _{q}e_{q}+(1-\alpha _{q})\sum_{\,r\neq q}\left( \max_{k\in
\mathcal{D}_{r}\cap \mathcal{D}_{q}}\left[ \mathbf{H}_{rq}\right]
_{kk}\right) e_{r},\qquad \forall \mathbf{p}^{(1)},\text{ }\mathbf{p}^{(2)}%
\mathbf{\in }{\mathscr{P}}^{\text{eff}}\text{,\quad }\forall q\in \Omega ,
\label{inequality-contraction_}
\end{align}%
where (\ref{Ineq_2}) follows from (\ref{Mapping_T}) and the triangle
inequality \cite{Horn85}; in (\ref{Ineq_3}) $\left\Vert \cdot \right\Vert
_{2,\mathcal{D}_{q}}$ is defined in (\ref{L2_Dq_norm}) and the equality
follows from the fact that ${\mathscr{P}}_{q}^{\text{eff}}$ sets to zero the
elements not in $\mathcal{D}_{q}$; (\ref{ineq_6}) follows from Lemma \ref%
{NonExpansive-Lemma} (see (\ref{NonExpansive_ineq})); in (\ref{ineq_7}) $%
\overline{\mathbf{H}}_{rq}$ is a diagonal matrix defined as%
\begin{equation}
\left[ \overline{\mathbf{H}}_{rq}\right] _{kk}%
\triangleq%
\left\{
\begin{array}{l}
\left[ \mathbf{H}_{rq}\right] _{kk}, \\
0,%
\end{array}%
\begin{array}{l}
\text{if }k\in \mathcal{D}_{r}\cap \mathcal{D}_{q}, \\
\text{otherwise.}%
\end{array}%
\right.   \label{def:H_bar_rq}
\end{equation}%
Introducing the vectors\vspace{-0.2cm}
\begin{equation}
\mathbf{e}_{\mathbf{T}}%
\triangleq%
[e_{\mathbf{T}_{1}},\ldots ,e_{\mathbf{T}_{Q}}]^{T},\mathbf{\quad }\text{and}%
\mathbf{\quad e}%
\triangleq%
[e_{1},\ldots ,e_{Q}]^{T},  \label{e_vec}
\end{equation}%
with $e_{\mathbf{T}_{q}}$ and$\ e_{q}$ defined in (\ref{e_q_and_e_T_q}), and
the matrix
\begin{equation}
\mathbf{H}_{\mathbf{\alpha }}^{\max }%
\triangleq%
\mathbf{D}_{\mathbf{\alpha }}+(\mathbf{I-D}_{\mathbf{\alpha }})\mathbf{H}%
^{\max },\text{ \ with \ }\mathbf{D}_{\mathbf{\alpha }}%
\triangleq%
\limfunc{diag}(\alpha _{q}\ldots \alpha _{Q}),
\end{equation}%
where $\mathbf{H}^{\max }$ is defined in (\ref{Hmax}). Then, the set of
inequalities in (\ref{inequality-contraction_}) for all $q,$ can be
rewritten in vectorial form as%
\begin{equation}
\mathbf{0}\leq \mathbf{e}_{\mathbf{T}}\leq \mathbf{H}_{\mathbf{\alpha }%
}^{\max }\mathbf{e},\quad \forall \mathbf{p}^{(1)},\text{ }\mathbf{p}^{(2)}%
\mathbf{\in }{\mathscr{P}}^{\text{eff}}\text{.}  \label{vector_e_}
\end{equation}%
Using the weighted maximum norm $\left\Vert \mathbf{\cdot }\right\Vert
_{\infty ,\text{vec}}^{\mathbf{w}}$ defined in (\ref%
{weighted_infinity_vector_norm}) in combination with (\ref{vector_e_}), we
have, $\forall \mathbf{p}^{(1)},$ $\mathbf{p}^{(2)}\mathbf{\in }{\mathscr{P}}%
^{\text{eff}}$ and\ $\forall \mathbf{w}>\mathbf{0,}$%
\begin{equation}
\left\Vert \mathbf{e}_{\mathbf{T}}\right\Vert _{\infty ,\text{vec}}^{\mathbf{%
w}}\leq \left\Vert \mathbf{H}_{\mathbf{\alpha }}^{\max }\mathbf{e}%
\right\Vert _{\infty ,\text{vec}}^{\mathbf{w}}\leq \left\Vert \mathbf{H}_{%
\mathbf{\alpha }}^{\max }\right\Vert _{\infty ,\text{mat}}^{\mathbf{w}%
}\left\Vert \mathbf{e}\right\Vert _{\infty ,\text{vec}}^{\mathbf{w}},
\label{e_t_contraction}
\end{equation}%
where $\left\Vert \cdot \right\Vert _{\infty ,\text{mat}}^{\mathbf{w}}$ is
the matrix norm induced by the vector norm $\left\Vert \mathbf{\cdot }%
\right\Vert _{\infty ,\text{vec}}^{\mathbf{w}}$ in (\ref%
{weighted_infinity_vector_norm}) and defined in (\ref{H_max_weight_norm})
\cite{Horn85}. Finally, using (\ref{e_t_contraction}) and (\ref%
{block_max_weight_norm}), we obtain, $\forall \mathbf{p}^{(1)},$ $\mathbf{p}%
^{(2)}\mathbf{\in }{\mathscr{P}}^{\text{eff}}$and\ $\forall \mathbf{w}>%
\mathbf{0},$
\begin{align}
\left\Vert \mathbf{T(\mathbf{p}}^{(1)}\mathbf{)}-\mathbf{T(\mathbf{p}}^{(2)}%
\mathbf{)}\right\Vert _{2,\text{block}}^{\mathbf{w}}& =\max_{q}\frac{%
\left\Vert \mathbf{T}_{q}\mathbf{(\mathbf{p}}^{(1)}\mathbf{)}-\mathbf{T}_{q}%
\mathbf{(\mathbf{p}}^{(2)}\mathbf{)}\right\Vert _{2}}{w_{q}}=\left\Vert
\mathbf{e}_{\mathbf{T}}\right\Vert _{\infty ,\text{vec}}^{\mathbf{w}}  \notag
\\
& \leq \left\Vert \mathbf{H}_{\mathbf{\alpha }}^{\max }\right\Vert _{\infty ,%
\text{mat}}^{\mathbf{w}}\left\Vert \mathbf{e}\right\Vert _{\infty ,\text{vec}%
}^{\mathbf{w}}=\left\Vert \mathbf{H}_{\mathbf{\alpha }}^{\max }\right\Vert
_{\infty ,\text{mat}}^{\mathbf{w}}\left\Vert \mathbf{\mathbf{p}}^{(1)}-%
\mathbf{\mathbf{p}}^{(2)}\right\Vert _{2,\text{block}}^{\mathbf{w}},
\end{align}%
which leads to a block-contraction for the mapping $\mathbf{T,}$ if $%
\left\Vert \mathbf{H}_{\mathbf{\alpha }}^{\max }\right\Vert _{\infty ,\text{%
mat}}^{\mathbf{w}}<1,$ implying condition (\ref{b_modulus_contraction})
(since each $\alpha _{q}\in \lbrack 0,1),$ $\forall q\in \Omega $).
\end{proof}

\section{Proof of Theorem \protect\ref{Theo-IWFA} and Theorem \protect\ref%
{Theo-IWFA_memory}}

\label{proof of Theorem Theo-IWFA}

Since the sequential IWFA described in Algorithm \ref{IWFA_Algo} is an
instance of the smoothed sequential IWFA given in Algorithm \ref%
{IWFA_Algo_MEMO} when $\alpha_q =0$ for all $q\in \Omega$, to prove convergence of both algorithms,
it is sufficient to show that Algorithm \ref{IWFA_Algo_MEMO}, under
condition (\ref{SF-IWFA}), \emph{globally} converges to the NE of game ${%
{\mathscr{G}}%
,}$ for\emph{\ any }given set  $\{\alpha_q\}_{q\in \Omega}$, provided that each $\alpha_q\in \lbrack 0,1).$ We thus focus in the
following only on Algorithm \ref{IWFA_Algo_MEMO}, w.l.o.g..

It follows from Corollary \ref{Corollary-Multiuser_WF_Projection} and (\ref%
{Mapping_T}) that Algorithm \ref{IWFA_Algo_MEMO} is just an instance of the
Gauss-Seidel scheme based on the mapping $\mathbf{T,}$ defined in (\ref%
{Mapping_T}). Observe that, to study the convergence of Algorithm \ref%
{IWFA_Algo_MEMO}, there is no loss of generality in considering the mapping $%
\mathbf{T}$ defined in ${\mathscr{P}}^{\text{eff}}\subset {\mathscr{P}}$
instead of ${\mathscr{P},}$ since all the points produced by the algorithm
(except possibly the initial point, which does not affect the convergence of
the algorithm in the subsequent iterations) as well as the Nash equilibria
of the game are confined, by definition, in ${\mathscr{P}}^{\text{eff}}$
(see (\ref{fixed-point_NE}) in Appendix \ref{Appendix_WF_project}).
Convergence of the Gauss-Seidel scheme based on the mapping $\mathbf{T}$ is
given by the following result that comes from {\cite[Prop. 1.4]{Bertsekas
Book-Parallel-Comp}}\footnote{%
Observe that the set ${\mathscr{P}}^{\text{eff}}$ defined in (\ref{X_q set})
is closed, as required in {\cite[Prop. 1.4]{Bertsekas Book-Parallel-Comp}}.}
and {\cite[Prop. 1.1a)]{Bertsekas Book-Parallel-Comp}}.

\begin{proposition}
\label{Bertsekas_proposition_Gauss-Seidel-mapping}If the mapping $\mathbf{T}:%
{\mathscr{P}}^{\text{eff}}\mathcal{\mapsto }{\mathscr{P}}^{\text{eff}}$
defined in (\ref{Mapping_T}) is a block-contraction with respect to some
vector norm, then: $1)$ The mapping $\mathbf{T}$ has a unique fixed point in
${\mathscr{P}}^{\text{eff}}$; $2)$ The sequence of vectors starting from any
arbitrary point in ${\mathscr{P}}^{\text{eff}}$ and generated by the
Gauss-Seidel algorithm based on the mapping $\mathbf{T}$, converges linearly
to the fixed point of $\mathbf{T}$.
\end{proposition}

It follows from Proposition \ref{Proposition:block-contraction} and Proposition \ref{Bertsekas_proposition_Gauss-Seidel-mapping} that the global
convergence of Algorithm \ref{IWFA_Algo_MEMO} is guaranteed under the
sufficient condition given in (\ref{b_modulus_contraction}). Moreover, since
(\ref{b_modulus_contraction}) does not depend on $\{\alpha\}_{q\in \Omega},$ the convergence
of the algorithm is not affected by the particular choice of $\alpha_q$'s as
well  [provided that each $\alpha_q \in \lbrack 0,1)$].

To complete the proof, we just need to show that (\ref{b_modulus_contraction}%
) is equivalent to (\ref{SF-IWFA}). \ Since $\mathbf{H}^{\max }$ is a
nonnegative matrix, there exists a positive vector $\overline{\mathbf{w}}$
such that \cite[Corollary 6.1]{Bertsekas Book-Parallel-Comp}%
\begin{equation}
\left\Vert \mathbf{H}^{\max }\right\Vert _{\infty ,\text{mat}}^{\overline{%
\mathbf{w}}}<1\quad \Leftrightarrow \quad \rho \left( \mathbf{H}^{\max
}\right) <1.\quad  \label{eq_nor_spectral_radius}
\end{equation}%
Since the convergence of Algorithm \ref{IWFA_Algo_MEMO} is guaranteed under (%
\ref{b_modulus_contraction}), for any given $\mathbf{w>0,}$ we can choose $%
\mathbf{w=}\overline{\mathbf{w}}$ and use (\ref{eq_nor_spectral_radius});
which proves the desired result.

Conditions (\ref{SF_for_C1_a})-(\ref{SF_for_C1_b}) in Corollary \ref%
{Corollary-C1-C2} can be obtained as follows. Using \cite[Proposition 6.2e]%
{Bertsekas Book-Parallel-Comp}%
\begin{equation}
\rho (\mathbf{H}^{\max })\leq \Vert \mathbf{H}^{\max }\Vert _{\infty ,\text{%
mat}}^{\mathbf{w}},\quad \forall \mathbf{w>0,}
\end{equation}%
a sufficient condition for the $\Rightarrow $ direction in (\ref%
{eq_nor_spectral_radius}) is%
\begin{equation}
\Vert \mathbf{H}^{\max }\Vert _{\infty ,\text{mat}}^{\mathbf{w}}<1,\quad
\end{equation}%
for some given $\mathbf{w>0;}$ which provides (\ref{SF_for_C1_a}). \ The
optimal vector $\mathbf{w}$ is given by the following Geometric Programming
\cite{Boyd}
\begin{equation*}
\begin{array}{ll}
\limfunc{minimize}\limits_{\mathbf{w}}\quad \max\limits_{q} &
\dsum\limits_{r=1,r\neq q}\left[ \mathbf{H}^{\max }\right]
_{qr}w_{q}^{-1}w_{r} \\
\text{subject to } & \mathbf{w}>\mathbf{0,}%
\end{array}%
\end{equation*}%
which provides (\ref{GP}).

Condition (\ref{SF_for_C1_b}) is obtained similarly, still using (\ref%
{eq_nor_spectral_radius}) and $\rho \left( \mathbf{H}^{\max }\right) =\rho
\left( \mathbf{H}^{\max T}\right) .$\hspace{\fill}\rule{1.5ex}{1.5ex}\vspace{%
-0.7cm}

\bigskip

\section{Proof of Theorem \protect\ref{Theo-SIWFA}}

\label{proof of Theorem Theo-SIWFA}The proof is based on the same steps as
in Appendix \ref{proof of Theorem Theo-IWFA}. It follows from \cite[Prop. 1.1%
]{Bertsekas Book-Parallel-Comp} that both Algorithm \ref{SIWFA_algo} and
Algorithm \ref{SIWFA_algo_MEMO} linearly converge to the unique NE of game ${%
{\mathscr{G}}%
,}$ {starting from any arbitrary point in }${\mathscr{P}}${,} if the mapping
$\mathbf{T}$ defined in (\ref{Mapping_T}) is a contraction (see (\ref%
{pseudo-contraction_def})) in some vector norm. Using the block-maximum norm
as defined in (\ref{block_max_weight_norm}), invoking Proposition \ref%
{Proposition:block-contraction}, and following the same approach as in
Appendix \ref{proof of Theorem Theo-IWFA} we obtain the desired sufficient
condition (\ref{SF-IWFA}) for the global convergence of both Algorithm 3 and Algorithm 4, for any given set $\{\alpha\}_{q\in \Omega},$ [provided that each $\alpha _{q}\in [0,1)]$.%
\hspace{\fill}\rule{1.5ex}{1.5ex}%

\section{Proof of Theorem \protect\ref{Theorem Theo-SIWFA_general}}

\label{proof of Theorem Theo-SIWFA_general} Since Algorithm \ref{SIWFA_algo}
is an instance of Algorithm \ref{SIWFA_algo_MEMO} when each $\alpha _{q}=0$,
we focus only on the latter, w.l.o.g.. The proof of the theorem is based on
(pseudo) contraction arguments, similarly to what we already shown in
Appendix \ref{proof of Theorem Theo-IWFA}. The main difference with respect
to the approach proposed in Appendix \ref{proof of Theorem Theo-IWFA} is due
to the alternative definition of the error vector generated by Algorithm \ref%
{SIWFA_algo_MEMO}, as detailed next.

Denoting by $\mathbf{p}^{(n)}%
\triangleq%
(\mathbf{p}_{q}^{(n)})_{q\in \Omega }$ the power allocation vector generated
by Algorithm \ref{SIWFA_algo_MEMO} at iteration $n\geq 1,$ \ with arbitrary
starting point $\mathbf{p}^{(0)}\in {\mathscr{P}},$ and using the mapping $%
\mathbf{T}$ defined in (\ref{Mapping_T}), we have
\begin{equation}
\mathbf{p}^{(n+1)}=\mathbf{T}(\mathbf{p}^{(n)}),\quad \forall n\geq 1.
\end{equation}%
Let $\mathbf{p}^{\star }%
\triangleq%
\mathbf{(\mathbf{p}_{q}^{\star })}_{q\in \Omega }$ be a NE of game ${%
{\mathscr{G}}%
}$ (and thus a fixed point of the mapping $\mathbf{T}$), whose existence is
guaranteed by \cite[Theorem 2]{Scutari-Part I}. Define the vector $\mathbf{e}%
^{(n)}%
\triangleq%
[\mathbf{e}_{1}^{(n)T},\ldots ,\mathbf{e}_{Q}^{(n)T}]^{T},$ with
\begin{equation}
\mathbf{e}_{q}^{(n)}%
\triangleq%
\mathbf{p}_{q}^{(n)}-\mathbf{p}_{q}^{\star },\quad n\geq 1\text{ and }q\in
\Omega ,  \label{total vector_e}
\end{equation}%
and, given $\{\alpha _{q}\}_{q\in \Omega }$ with $\alpha _{q}\in \lbrack
0,1),$ define the $Q\times Q$ matrix $\mathbf{D}_{\mathbf{\alpha }}%
\triangleq%
\limfunc{diag}(\alpha _{1}\ldots \alpha _{Q}).$ Then, for each $n\geq 1,$ we
have%
\begin{eqnarray}
\hspace{-1cm}\left\Vert \mathbf{e}^{(n+1)}\right\Vert _{2} &\leq &\left\Vert
(\mathbf{D}_{\mathbf{\alpha }}\otimes \mathbf{I}_{N})\mathbf{e}%
^{(n)}\right\Vert _{2}+\left\Vert \left[
\begin{array}{c}
-(1-\alpha _{1})\left( \mathbf{\sigma }_{1}+\sum_{\,r\neq 1}\mathbf{H}_{r1}%
\mathbf{p}_{r}^{(n)}\right)  \\
\vdots  \\
-(1-\alpha _{Q})\left( \mathbf{\sigma }_{Q}+\sum_{\,r\neq Q}\mathbf{H}_{rQ}%
\mathbf{p}_{r}^{(n)}\right)
\end{array}%
\right] _{{\mathscr{P}}^{\text{eff}}}\right.   \notag \\
&&\left. -\left[
\begin{array}{c}
-(1-\alpha _{1})\left( \mathbf{\sigma }_{1}+\sum_{\,r\neq 1}\mathbf{H}_{r1}%
\mathbf{p}_{r}^{\star }\right)  \\
\vdots  \\
-(1-\alpha _{Q})\left( \mathbf{\sigma }_{Q}+\sum_{\,r\neq Q}\mathbf{H}_{rQ}%
\mathbf{p}_{r}^{\star }\right)
\end{array}%
\right] _{{\mathscr{P}}^{\text{eff}}}\right\Vert _{2}  \label{Set_ineq_2} \\
&\leq &\left\Vert (\mathbf{D}_{\mathbf{\alpha }}\otimes \mathbf{I}%
_{N})\right\Vert _{2,\text{mat}}\left\Vert \mathbf{e}^{(n)}\right\Vert
_{2}+\left\Vert \left[
\begin{array}{c}
(1-\alpha _{1})\sum_{\,r\neq 1}\overline{\mathbf{H}}_{r1}\left( \mathbf{p}%
_{r}^{(n)}-\mathbf{p}_{r}^{\star }\right)  \\
\vdots  \\
(1-\alpha _{Q})\sum_{\,r\neq Q}\overline{\mathbf{H}}_{rQ}\left( \mathbf{p}%
_{r}^{(n)}-\mathbf{p}_{r}^{\star }\right)
\end{array}%
\right] \right\Vert _{2}  \label{Set_ineq_3} \\
&=&\left\Vert (\mathbf{D}_{\mathbf{\alpha }}\otimes \mathbf{I}%
_{N})\right\Vert _{2,\text{mat}}\left\Vert \mathbf{e}^{(n)}\right\Vert
_{2}+\left\Vert (\mathbf{I}_{N}\otimes (\mathbf{I}_{Q}-\mathbf{D}_{\mathbf{%
\alpha }}))\mathbf{HPe}^{(n)}\right\Vert _{2}  \label{Set_ineq_4} \\
&\leq &\left( \left\Vert (\mathbf{D}_{\mathbf{\alpha }}\otimes \mathbf{I}%
_{N})\right\Vert _{2,\text{mat}}+\left\Vert (\mathbf{I}_{N}\otimes (\mathbf{I%
}_{Q}-\mathbf{D}_{\mathbf{\alpha }})\right\Vert _{2,\text{mat}}\left\Vert
\mathbf{H}\right\Vert _{2,\text{mat}}\right) ^{n}\left\Vert \mathbf{e}%
^{(1)}\right\Vert _{2},  \label{Set_ineq_last}
\end{eqnarray}%
where (\ref{Set_ineq_2}) follows from the triangle inequality and ${%
\mathscr{P}}^{\text{eff}}={\mathscr{P}}_{1}^{\text{eff}}\times \cdots \times
{\mathscr{P}}_{Q}^{\text{eff}},$ where \textquotedblleft $\otimes $%
\textquotedblright\ denotes the Kronecker product$;$ (\ref{Set_ineq_3})
follows from the nonexpansive property of the Euclidean projector, the
definition of ${\mathscr{P}}^{\text{eff}}$ (see (\ref{X_q set})) and the
definition of the diagonal matrices $\overline{\mathbf{H}}_{rq},$ as given
in (\ref{def:H_bar_rq}); and in (\ref{Set_ineq_4}) we have used a
permutation matrix $\mathbf{P}$ so that the vector $\mathbf{e}^{(n)}$ given
in (\ref{total vector_e}), is replaced by $\widetilde{\mathbf{e}}^{(n)}%
\triangleq%
\mathbf{Pe}^{(n)}=[\widetilde{\mathbf{e}}_{1}^{(n)T},\ldots ,\widetilde{%
\mathbf{e}}_{N}^{(n)T}]^{T},$ with $\widetilde{\mathbf{e}}_{k}^{(n)}%
\triangleq%
[p_{1}^{(n)}(k),\ldots ,p_{Q}^{(n)}(k)]^{T}-[p_{1}^{\star }(k),\ldots
,p_{Q}^{\star }(k)]^{T},$ and the matrix $\mathbf{H}$ is defined as%
\begin{equation}
\mathbf{H}%
\triangleq%
\limfunc{diag}\left( \mathbf{H(}1\mathbf{),\ldots ,H(}N\mathbf{)}\right) ,
\label{def:H}
\end{equation}%
with $\mathbf{H(}k\mathbf{)}$ given in (\ref{H(k)_qr}). \ The matrix norm $%
\left\Vert \mathbf{H}\right\Vert _{2,\text{mat}}$ in (\ref{Set_ineq_3}) is
the spectral norm (induced by the vector Euclidean norm \cite{Horn85}),
defined as $\left\Vert \mathbf{H}\right\Vert _{2,\text{mat}}%
\triangleq%
\rho ^{1/2}\left( \mathbf{H}^{T}\mathbf{H}\right) .$

From (\ref{Set_ineq_last}) it follows that Algorithm \ref{SIWFA_algo_MEMO}
converges to the NE $\mathbf{p}^{\star },$ from any starting point $\mathbf{p%
}^{(0)}\in {\mathscr{P},}$ if $\left( \left\Vert (\mathbf{D}_{\mathbf{\alpha
}}\otimes \mathbf{I}_{N})\right\Vert _{2,\text{mat}}+\left\Vert (\mathbf{I}%
_{N}\otimes (\mathbf{I}_{Q}-\mathbf{D}_{\mathbf{\alpha }})\right\Vert _{2,%
\text{mat}}\left\Vert \mathbf{H}\right\Vert _{2,\text{mat}}\right) ^{n}$ in (%
\ref{Set_ineq_last}) approaches to zero as $n\rightarrow \infty ,$ which is
guaranteed if the following conditions are satisfied%
\begin{equation*}
\rho ^{1/2}\left( \mathbf{H}^{T}(k)\mathbf{H(}k\mathbf{)}\right) <\frac{%
1-\max_{q\in \Omega }\alpha _{q}}{1-\min_{q\in \Omega }\alpha _{q}},\quad
\forall k\in \{1,\ldots ,N\},
\end{equation*}%
which provides the desired result. Given (\ref{Set_ineq_last}), the linear
convergence of the algorithm follows directly from \cite[Sec. 1.3.1]%
{Bertsekas Book-Parallel-Comp}.
\hspace{\fill}\rule{1.5ex}{1.5ex}%
\vspace{-0.4cm}

\section{Proof of Convergence of Algorithm \protect\ref{SGPA_algo} and
Algorithm \protect\ref{GPA_algo}}

\label{proof Theorem Theo-SGPA}The global convergence of both sequential and
simultaneous IGPAs, described in Algorithms \ref{SGPA_algo} and \ref%
{GPA_algo}, is guaranteed if Algorithms \ref{SGPA_algo} and \ref{GPA_algo}
satisfy {\cite[Prop. 1.1]{Bertsekas Book-Parallel-Comp}} and {\cite[Prop. 1.4%
]{Bertsekas Book-Parallel-Comp}}, respectively. To this end, since each ${%
\mathscr{P}}_{q}$ is compact (and thus also ${\mathscr{P}=\mathscr{P}}%
_{1}\times \cdots \times {\mathscr{P}}_{Q})$, it is sufficient that the
mapping $\mathbf{T}_{\mathbf{G}}(\mathbf{p})=\mathbf{(T}_{\mathbf{G}%
_{q}}^{T}(\mathbf{p}))_{q\in \Omega }:{\mathscr{P}\mapsto \mathscr{P}}$ is a
block-contraction (see (\ref{pseudo-contraction_def})) with respect to the
norm $\left\Vert \cdot \right\Vert _{\mathbf{G},\text{block}},$ defined as
\begin{equation}
\left\Vert \mathbf{T}_{\mathbf{G}}(\mathbf{p})\right\Vert _{\mathbf{G},\text{%
block}}%
\triangleq%
\max_{q\in \Omega }\left\Vert \mathbf{T}_{\mathbf{G}_{q}}(\mathbf{p}%
)\right\Vert _{\mathbf{G}_{q},2},\quad \text{with}\quad \left\Vert \mathbf{T}%
_{\mathbf{G}_{q}}(\mathbf{p})\right\Vert _{\mathbf{G}_{q},2}%
\triangleq%
(\mathbf{T}_{\mathbf{G}_{q}}^{T}(\mathbf{p})\mathbf{G}_{q}\mathbf{T}_{%
\mathbf{G}_{q}}(\mathbf{p}))^{1/2},  \label{block-maximum}
\end{equation}%
{where} $\mathbf{T}_{\mathbf{G}_{q}}(\mathbf{p})$\ is defined as%
\begin{equation}
\mathbf{T}_{\mathbf{G}q}(\mathbf{p})%
\triangleq%
\left[ \mathbf{p}_{q}-\beta \mathbf{G}_{q}^{-1}f_{q}(\mathbf{p}_{q},\mathbf{p%
}_{-q})\right] _{{\mathscr{P}}_{q}}^{\mathbf{G}_{q}},\quad \text{with}\quad
f_{q}(\mathbf{p}_{q},\mathbf{p}_{-q})%
\triangleq%
-%
\nabla%
_{q}R_{q}(\mathbf{p}_{q},\mathbf{p}_{-q}).  \label{Mapping_T_G}
\end{equation}%
Rewriting $\mathbf{T}_{\mathbf{G}_{q}}(\mathbf{p})$ in (\ref{Mapping_T_G})
as
\begin{equation}
\mathbf{T}_{\mathbf{G}q}(\mathbf{p})%
\triangleq%
\left[ \mathbf{R}_{\mathbf{G}q}(\mathbf{p})\right] _{{\mathscr{P}}_{q}}^{%
\mathbf{G}_{q}},\text{ \ \ with \ }\mathbf{R}_{\mathbf{G}_{q}}(\mathbf{p})%
\triangleq%
\mathbf{p}_{q}-\beta \mathbf{G}_{q}^{-1}f_{q}(\mathbf{p}_{q},\mathbf{p}%
_{-q}),  \label{T_g as_R_G_def}
\end{equation}%
and using (\ref{block-maximum}), we obtain: $\forall \mathbf{p}^{(1)},%
\mathbf{p}^{(2)}\in {\mathscr{P},}$%
\begin{eqnarray}
\left\Vert \mathbf{T}_{\mathbf{G}}(\mathbf{p}^{(1)})-\mathbf{T}_{\mathbf{G}}(%
\mathbf{p}^{\left( 2\right) })\right\Vert _{\mathbf{G},\text{block}}
&=&\max_{q}\left\Vert \left[ \mathbf{R}_{\mathbf{G}q}(\mathbf{p}^{(1)})%
\right] _{{\mathscr{P}}_{q}}^{\mathbf{G}_{q}}-\left[ \mathbf{R}_{\mathbf{G}%
q}(\mathbf{p}^{\left( 2\right) })\right] _{{\mathscr{P}}_{q}}^{\mathbf{G}%
_{q}}\right\Vert _{\mathbf{G}_{q},2}  \notag \\
&\leq &\max_{q}\left\Vert \mathbf{R}_{\mathbf{G}q}(\mathbf{p}^{(1)})-\mathbf{%
R}_{\mathbf{G}q}(\mathbf{p}^{\left( 2\right) })\right\Vert _{\mathbf{G}%
_{q},2}=\left\Vert \mathbf{R}_{\mathbf{G}}(\mathbf{p}^{(1)})-\mathbf{R}_{%
\mathbf{G}}(\mathbf{p}^{\left( 2\right) })\right\Vert _{\mathbf{G},\text{%
block}},  \label{contraction on R_G}
\end{eqnarray}%
where the inequality follows from the non-expansive property of the
projection $\left[ \mathbf{\cdot }\right] _{{\mathscr{P}}_{q}}^{\mathbf{G}%
_{q}}$ in the norm $\left\Vert \mathbf{\cdot }\right\Vert _{\mathbf{G}%
_{q},2} $. From (\ref{contraction on R_G}) it follows that a sufficient
condition for $\mathbf{T}_{\mathbf{G}}$ being a contraction with respect to
the norm $\left\Vert \cdot \right\Vert _{\mathbf{G},\text{block}}$ in (\ref%
{block-maximum}) is that the mapping $\mathbf{R}_{\mathbf{G}}(\mathbf{p})%
\triangleq%
(\mathbf{R}_{\mathbf{G}_{1}}(\mathbf{p}))_{q\in \Omega }:{\mathscr{P}}%
\mapsto
\mathbb{R}
^{QN}$ defined in (\ref{T_g as_R_G_def}) be a contraction with respect to
the same norm.

We derive now sufficient conditions for $\mathbf{R}_{\mathbf{G}}(\mathbf{p})$
being a contraction with respect to $\left\Vert \mathbf{\cdot }\right\Vert _{%
\mathbf{G},\text{block}}$ defined in (\ref{block-maximum}). For the sake of
simplicity, we will consider only the case in which $\mathbf{G}_{q}=\mathbf{%
I,}$ $\forall q\in \Omega .$

We introduce the following notation: For any $f_{q}(\mathbf{p})%
\triangleq%
-%
\nabla%
_{q}R_{q}(\mathbf{p}_{q},\mathbf{p}_{-q}),$ let $%
\nabla%
_{r}f_{q}(\mathbf{p})$ denote the $N\times N$ matrix, whose $j$-th column is
the gradient vector of the $j$-th component of $f_{q}(\mathbf{p}),$ when
viewed as function of $\mathbf{p}_{r}.$ Then, we have the following result
that comes directly from {\cite[Prop. 1.10]{Bertsekas Book-Parallel-Comp}}.

\begin{proposition}
As \textrm{Nit}$\longrightarrow \infty ,$ the IGPAs described in Algorithms %
\ref{SGPA_algo} and \ref{GPA_algo} converge to the unique NE of game ${%
{\mathscr{G}}%
}$ from any set of initial conditions in ${\mathscr{P},}$ if there exists a
scalar $\delta \in \lbrack 0,1)$ such that
\begin{equation}
\left\Vert \mathbf{I-}\beta
\nabla%
_{q}f_{q}(\mathbf{p})\right\Vert _{2,\text{mat}}+\dsum\limits_{r\neq
q}\left\Vert \beta
\nabla%
_{r}f_{q}(\mathbf{p})\right\Vert _{2,\text{mat}}\leq \delta ,\quad \forall
\mathbf{p}\in {\mathscr{P},}\text{ }\forall q\in \Omega ,  \label{Cond_1}
\end{equation}%
where $\left\Vert \mathbf{A}\right\Vert _{2,\text{mat}}$ denotes the
spectral norm of the matrix $\mathbf{A.}$
\end{proposition}

We derive now a sufficient condition for (\ref{Cond_1}). Using $\ f_{q}(%
\mathbf{p})%
\triangleq%
-%
\nabla%
_{q}R_{q}(\mathbf{p}_{q},\mathbf{p}_{-q})$ and (\ref{Rate}) we have%
\begin{equation}
\nabla%
_{r}f_{q}(\mathbf{p})=\mathbf{D}_{q}(\mathbf{p})\mathbf{H}_{rq},
\label{Nabla_matrix}
\end{equation}%
where%
\begin{eqnarray}
{\mathbf{H}_{rq}}\triangleq \limfunc{diag}\left( \left\{ \frac{%
|H_{rq}(k)|^{2}}{|H_{qq}(k)|^{2}}\right\} _{k}\right) ,\quad \mathbf{D}_{q}(%
\mathbf{p}) &\triangleq &\limfunc{diag}\left( \left\{ \dfrac{1}{\left(
\dfrac{1}{|H_{qq}(k)|^{2}}+\sum\limits_{r=1}^{Q}\Gamma _{q}^{-\delta _{rq}}%
\dfrac{|H_{rq}(k)|^{2}}{|H_{qq}(k)|^{2}}p_{r}(k)\right) ^{2}}\right\}
_{k}\right) .  \notag  \label{Dq} \\
&&
\end{eqnarray}%
Using (\ref{Dq}), condition (\ref{Cond_1}) becomes%
\begin{equation}
\max_{k}\left\vert 1\mathbf{-}\beta \left[ \mathbf{D}_{q}(\mathbf{p})\right]
_{kk}\right\vert +\beta \dsum\limits_{r\neq q}\max_{k}\left[ \mathbf{D}_{q}(%
\mathbf{p})\mathbf{H}_{rq}\right] _{kk}\leq \delta ,\quad \forall \mathbf{p}%
\in {\mathscr{P},}\text{ }\forall q\in \Omega .  \label{cond_4}
\end{equation}%
A sufficient condition for (\ref{cond_4}) is%
\begin{equation}
\dsum\limits_{r\neq q}\max_{k}\left[ \mathbf{H}_{rq}\right] _{kk}\leq \frac{%
\delta -\max_{k}\left\vert 1\mathbf{-}\beta \left[ \mathbf{D}_{q}(\mathbf{p})%
\right] _{kk}\right\vert }{\beta \max_{k}\left[ \mathbf{D}_{q}(\mathbf{p})%
\right] _{kk}},\quad \forall \mathbf{p}\in {\mathscr{P},}\text{ }\forall
q\in \Omega .  \label{Cond_5}
\end{equation}%
It is straightforward to see that it is always possible to find proper
(sufficiently small) $\beta >0$ and $\delta \in \lbrack 0,1)$ (close to one)
such that (\ref{Cond_5}) is satisfied, provided that
\begin{equation}
\dsum\limits_{r\neq q}\max_{k}\left[ \mathbf{H}_{rq}\right] _{kk}=\Gamma
_{q}\dsum\limits_{r\neq q}\max_{k}\frac{\left\vert \bar{H}%
_{rq}(k)\right\vert ^{2}}{\left\vert \bar{H}_{qq}(k)\right\vert ^{2}}\dfrac{%
d_{qq}^{\gamma }}{d_{rq}^{\gamma }}\dfrac{P_{r}}{P_{q}}<\varepsilon
_{q},\quad \forall q\in \Omega ,
\end{equation}%
where%
\begin{equation}
\varepsilon _{q}=\min_{\mathbf{p}\in {\mathscr{P}}}\frac{\min_{k}\left[
\mathbf{D}_{q}(\mathbf{p})\right] _{kk}}{\max_{k}\left[ \mathbf{D}_{q}(%
\mathbf{p})\right] _{kk}}\leq 1,
\end{equation}
and $\mathbf{D}_{q}(\mathbf{p})$ is defined in (\ref{Dq}). \hspace{\fill}%
\rule{1.5ex}{1.5ex}

\def\baselinestretch{0.2}

\end{document}